\documentclass[aps,prl,numerical,showkeys,showpacs,superscriptaddress,amsmath,amssymb,floatfix,preprint]{revtex4-1}

\usepackage[utf8]{inputenc}
\usepackage{subcaption}
\usepackage{amsmath}
\usepackage{amssymb}
\usepackage{lmodern}
\usepackage[pdftex]{graphicx}

\usepackage{xcolor}
\definecolor{purple}{RGB}{148,103,189}
\definecolor{cyan}{RGB}{0,191,191}
\definecolor{peach}{RGB}{253,192,134}
\definecolor{basinblue}{RGB}{56,108,176}
\definecolor{basingrey}{RGB}{102,102,102}
\definecolor{basingreen}{RGB}{127,201,127}

\usepackage{tikz}
\DeclareRobustCommand\full  {\tikz[baseline=-0.6ex]\draw[red,thick] (0,0)--(0.5,0);}

\DeclareRobustCommand\dashed{\tikz[baseline=-0.6ex]\draw[purple,thick,dashed] (0,0)--(0.54,0);}
\DeclareRobustCommand\chain {\tikz[baseline=-0.6ex]\draw[blue,thick,dash dot] (0,0)--(0.5,0);}
\DeclareRobustCommand\inf  {\tikz[baseline=-0.6ex]\draw[cyan,thick] (0,0)--(0.5,0);}
\DeclareRobustCommand\infzero {\tikz[baseline=-0.6ex]\draw[cyan,thick,dash dot] (0,0)--(0.5,0);}
\DeclareRobustCommand\cont  {\tikz[baseline=-0.6ex]\draw[purple,thick] (0,0)--(0.5,0);}
\DeclareRobustCommand\sync {\tikz[baseline=-0.6ex]\fill[peach] (-0.1,-0.1) rectangle (0.1,0.1);}
\DeclareRobustCommand\sol {\tikz[baseline=-0.6ex]\fill[basinblue] (-0.1,-0.1) rectangle (0.1,0.1);}
\DeclareRobustCommand\exot {\tikz[baseline=-0.6ex]\fill[basingreen] (-0.1,-0.1) rectangle (0.1,0.1);}
\DeclareRobustCommand\other {\tikz[baseline=-0.6ex]\fill[basingrey] (-0.1,-0.1) rectangle (0.1,0.1);}

\usepackage{hyperref}

\begin{document}

\title{Network-induced multistability: Lossy coupling and exotic solitary states}

\author{Frank \surname{Hellmann}}
\email{hellmann@pik-potsdam.de}
\author{Paul \surname{Schultz}}
\affiliation{Potsdam Institute for Climate Impact Research (PIK), Member of the Leibniz Association, P.O. Box 60 12 03, D-14412 Potsdam, Germany}
\author{Patrycja \surname{Jaros}}
\affiliation{Division of Dynamics, Łódź University of Technology,Stefanowskiego 1/15, 90-924 Łódź, Poland}
\author{Roman \surname{Levchenko}}
\affiliation{Faculty of Radiophysics, Electronics and Computer Systems, Taras Shevchenko National University of Kyiv, Volodymyrska St. 60, 01030 Kyiv, Ukraine}
\author{Tomasz \surname{Kapitaniak}}
\affiliation{Division of Dynamics, Łódź University of Technology,Stefanowskiego 1/15, 90-924 Łódź, Poland}
\author{J\"{u}rgen \surname{Kurths}}
\affiliation{Potsdam Institute for Climate Impact Research (PIK), Member of the Leibniz Association, P.O. Box 60 12 03, D-14412 Potsdam, Germany}
\affiliation{Department of Physics, Humboldt University of Berlin, Newtonstr. 15, 12489 Berlin, Germany}
\affiliation{Saratov State University, Saratov, Russia}
\author{Yuri \surname{Maistrenko}}
\affiliation{Potsdam Institute for Climate Impact Research (PIK), Member of the Leibniz Association, P.O. Box 60 12 03, D-14412 Potsdam, Germany}
\affiliation{Institute of Mathematics and Centre for Medical and Biotechnical Research, National Academy of Sciences of Ukraine, Tereshchenkivska St. 3, 01030 Kyiv, Ukraine}

\begin{abstract}  
The stability of synchronised networked systems is a multi-faceted challenge for
many natural and technological fields, from cardiac and neuronal tissue pacemakers to power grids. In the latter case, the ongoing transition to
distributed renewable energy sources is leading to a proliferation of dynamical actors. The
desynchronization of a few or even one of those would likely result in a substantial blackout. Thus
the dynamical stability of the synchronous state has become a focus of power grid research in recent
years.

In this letter we uncover that the non-linear stability against large perturbations is dominated and
threatened by the presence of \textit{solitary states} in 
which individual actors desynchronise.  Remarkably, when taking physical losses in the network into
account, the back-reaction of the network induces new {\it exotic} solitary states in the individual
actors, and the stability characteristics of the synchronous state are dramatically altered. These
novel effects will have to be explicitly taken into account in the design of future power grids, and
their existence poses a challenge for control.

While this letter focuses on power grids, the form of the coupling we explore here is generic, and
the presence of new states is very robust. We thus strongly expect the results presented here to
transfer to other systems of coupled heterogeneous Newtonian oscillators.  
\end{abstract}

\maketitle

The power grid is a vast network connecting generators and consumers of electrical energy. Due to the ongoing energy
transition, dynamical actors are becoming more numerous and heterogeneous, and new dynamical phenomena are expected to
occur in future power grids. This has brought the dynamical stability of the necessary 50/60Hz synchronous state
into sharp focus, and spurred a large number of theoretical works on this topic recently 
\cite{dorfler2013synchronization,dorfler2012synchronization,motter2013spontaneous,schiffer2014conditions,
schiffer2016survey,rohden2012self,rohden2014impact,witthaut2012braess}. 
It is known that these oscillator networks can be multistable. Strong perturbations can
move the power grid dynamics out of the basin of attraction of the synchronous state
\cite{menck2013basin,menck2014dead,hellmann2016survivability,mitra2017multiple}: synchrony collapses and blackout is
the likely result.

The question of the stability of synchronisation is not specific to power grids, but is central to a wide range of
systems, like coupled Josephson junctions and laser systems \cite{larger2015laser,wiesenfeld1998frequency}, animal and
bacterial flocking behaviour \cite{vicsek2012collective,chen2017weak}, Huygens's pendulum clocks
\cite{bennett2002huygens}, crowd synchrony on London Millennium Bridge \cite{Abdulrehem2009,belykh2017foot} and 
chemical \cite{Tinsley2012,Nkomo2013} or mechanical oscillators \cite{Martens2013,Kapitaniak2015}, as well as 
networks of neurons \cite{herzog2007neurons}.

In this letter we show that the non-linear dynamics of oscillators in power grids is fundamentally altered by the
resistance of the lines (i.e. transfer conductances) and the resulting energy losses. 
The line resitances translate to a phase-lagged coupling as in the Kuramoto-Sakaguchi 
model \cite{Sakaguchi1986b} which is a characteristic of most systems that can be described as coupled 
phase oscillators, e.g. by using a phase reduction approach \cite{Nakao2016}. 
The presence of losses hence breaks the coupling symmetry and hampers a rigorous mathematical
analysis, e.g. in terms of Lyapunov functions \cite{Chiang1989,Tan1993}.
As a trade-off in favour of analysability, models are typically studied assuming a 
lossless regime, while less is known about the lossy case (e.g. \cite{dorfler2012synchronization,dorfler2013synchronization,Vu2015}).

In the following, we address how a lagged coupling alters the multistability of the system with 
a focus on possible pathways of desynchronisation.
In particular we find that the presence of losses can vastly increase the
likelihood that a random perturbation hits the basin of attraction of a
solitary state. Some parameter regimes solely exhibiting synchronisation without losses
become  multistable in the presence of even minimal losses. We
uncover novel solitary states with large basins that are qualitatively
different to known, previously studied multistable states of the power grid.
Remarkably, these effects appear in all parameter regimes and topologies studied.
They seem to be completely robust, not relying on underlying symmetries.
 
It follows that, contrary to common belief in the community, 
losses are more important to be considered than expected. 
Purely network properties can fundamentally
alter the dynamics of the oscillators in the system. 
Hence developing a detailed understanding of the non-linear network
physics of power grids is of great practical importance going forward. 

The research into the stability of the synchronous states of power grids in the theoretical physics and control engineering community is
mainly performed using the Kuramoto model with inertia, also known as the swing equation, in which the phase $\phi_i$ of the
complex voltage at the nodes is the dynamical variable. From engineering practice it is known that this equation
accurately captures the short time behaviour of synchronous machines used today \cite{Weckesser2013,Auer2016}, 
but it also serves as a fairly general
starting point for the study of future control dynamics \cite{schiffer2016survey}. It accounts for a proportional
response of the frequency to power imbalance and the presence of inertia, but neglects higher order internal dynamics 
as well as variations in the voltage magnitude. In the reference frame co-rotating with the synchronous frequency,
the equations for $n$ nodes are given as:

\begin{align}
\label{eq:power grid}
H_i \ddot \phi_i &= P_i - D_i \dot \phi_i - \sum\limits_{j=1}^n P_{ij}\;,\\
P_{ij} &= K_{ij} \Biglb( \sin\left(\alpha_{ij}\right) + \sin\left( \phi_i - \phi_j - \alpha_{ij}\right)\Bigrb)\;.
\end{align}

Here $P_i$ is the power injected/consumed at node $i$, $D_i$ characterizes the power's response to frequency changes,
$H_{i}$ the inertia present and $P_{ij}$ is the power injected at node $i$ into the line connecting nodes $i$ and $j$.
This power is given in terms of the phase difference and the complex admittance $Y_{ij} = - i K_{ij} \exp(i\alpha_{ij})$, 
with $K$ and $\alpha$ typically positive (see derivation in SI). This is the inverse of the complex impedance
consisting of the reactance $X$ and the resistance $R$: $Y_{ij} = \frac{1}{R_{ij} + i X_{ij}}$. All quantities are in
the per unit system, such that the voltage magnitude equals one. Typical parameter choices are discussed in the Methods
section.

\begin{figure}[!ht]
  \centering
   \begin{subfigure}[t]{0.48\columnwidth}
    \caption{}
    \includegraphics[width=\textwidth]{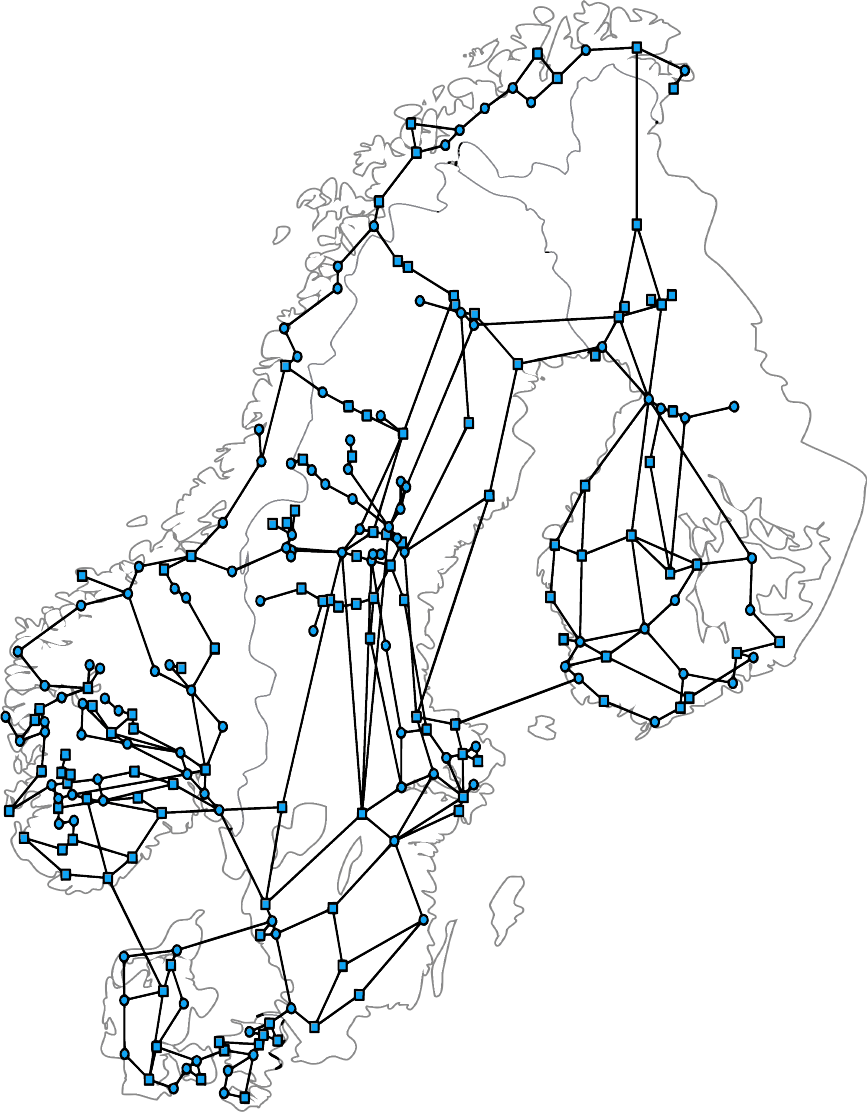}
    \label{fig1a}
  \end{subfigure}
   \begin{subfigure}[t]{0.48\columnwidth}
    \caption{}
    \includegraphics[width=.7\textwidth]{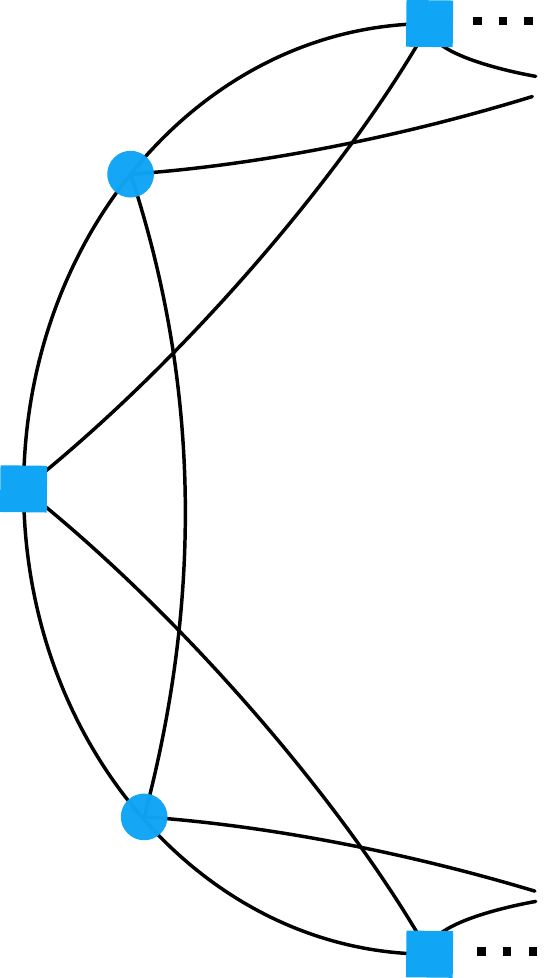}
    \label{fig1b}
  \end{subfigure}
  \caption{\textbf{Network models.} \textbf{a} The Scandinavian (extra-)high voltage transmission grid. 
  \textbf{b} A circle topology with coupling to next-nearest neighbours, i.e. a coupling radius of $R=2$.
  In both cases, squares denote net consumers and circles net producers.}
 \label{fig1}
\end{figure}

As discussed above, many authors follow the lossless assumption and assume that 
transmission lines are purely inductive.
 That is, $R_{ij} = 0$ and thus $\alpha_{ij} =
\arctan\left(\frac{R_{ij}}{X_{ij}}\right) = 0$. While it is well known to what
degree considering more realistic models of non-linear oscillators on the
nodes changes the overall picture, it is generally assumed that neglecting
losses does not have a major influence on the stability properties of the
dynamics.

In the engineering literature, the standard analysis of the return to
synchrony after a frequency event at a node neglects the back-reaction of the
dynamics at node $i$ on the other nodes and keeps them fixed, leading to the
infinite bus model:

\begin{align} 
\label{eq:infinite-bus-model}
 H \ddot \phi &= P - D \dot \phi + K \sin(\alpha) - K \sin(\phi - \alpha)\;.
\end{align}

When decoupling this system ($K = 0$) the oscillator rotates freely with
frequency $\omega_\text{lc} = \frac{P}{D}$. When the coupling is switched on
this limit cycle persists, and in the absence of losses it's average frequency
stays close to $\frac{P}{D}$\cite{menck2014dead}. This can be
seen as a simple model for solitary states \cite{Maistrenko2014, jaros2018solitary}, where the
infinite bus represents the remaining synchronous component.

\begin{figure*}[!ht]

\begin{subfigure}[b]{\textwidth}
    \caption{Scandinavian grid}
    \includegraphics[width=\textwidth,trim={0 0 22cm 0},clip]{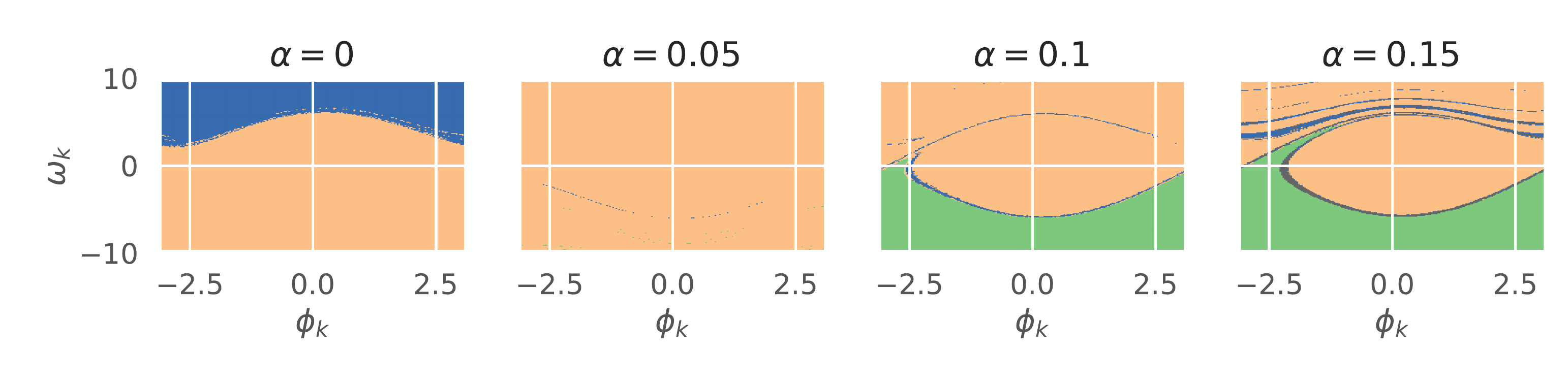}
    \label{fig2a}
  \end{subfigure}

  \begin{subfigure}[b]{\textwidth} 
    \caption{Circle topology}
    \includegraphics[width=\textwidth,trim={0 0 22cm 0},clip]{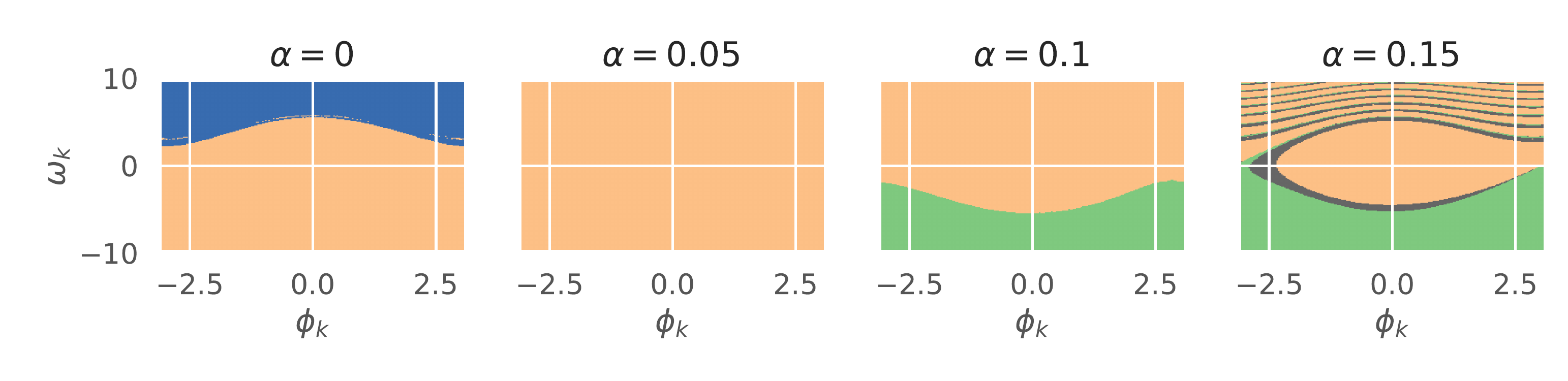}
    \label{fig2b}
  \end{subfigure} 
  \caption{\textbf{Phase space cross sections.}
  Cross section of the phase space corresponding to phase $\phi_k$ and phase velocity $\omega_k$ of a
  randomly chosen node of \textbf{a} the Scandinavian power grid and \textbf{b} the circle
  topology (both with standard parametrisation and control, see Methods). Each point belongs to the sync basin (\sync), the basin of a solitary state rotating 
  naturally (\sol) or in the basin of an exotic solitary state (\exot). Other asymptotic states are marked in grey (\other).
  Further parametrisations are given in SI.
  }
  \label{fig2}
\end{figure*}

As we will see, the physics of the power lines has a major impact on the non-linear dynamics of the
power grid. In the presence of losses, the average frequency of the limit cycle is moved to 
$\omega_\text{lc}^\prime = \omega_{lc} - \frac{K}{D}\sin(\alpha)$. 
This shift dramatically alters the basin structure of the solitary, and can even
flip the sign of the rotation. This leads to \emph{exotic solitary states}, where an oscillator rotates in the opposite
direction when coupled to the system than when uncoupled, even though the rest of the system remains in synchrony.

The dynamic effects due to losses are considerably more important than a variety of modelling assumptions that have been studied in the
literature until now. They are already very pronounced for currently most studied high voltage power grids
($\alpha_{ij} \approx 0.24$, see Methods section) and become a dominant factor for future decentralized energy
production with much higher losses on medium voltage power lines ($\alpha_{ij} \approx 1.4$), 
which are close to the theoretical stability limit of $\frac{\pi}{2}$ in the lossless case.

We begin by studying the phase space slice corresponding to the degrees of freedom
of a randomly chosen node. The concrete systems we consider are the well-studied
Scandinavian power grid (Fig.~\ref{fig1a}) with $n=236$ nodes  and a mean
degree of $\bar{d}=2.7$ \cite{menck2014dead,schultz2014detours} as well as a
regular circle topology (Fig.~\ref{fig1b}) with $n=50$ nodes, a coupling
radius of $R=2$ and otherwise equivalent parameter choices (see Methods
section for details on the parametrization and SI for extensive meanfield analysis of this model). 
In the Scandinavian topology, an equal number of
net producers and consumers is randomly distributed, while they are placed
alternately on the circle topology.

Fig.~\ref{fig2} shows the result for both networks. The color indicates the asymptotic state
reached after running the system with the chosen node in the plotted initial
condition, and the other nodes in the synchronous state. We immediately see a dramatic change in structure for the range of $\alpha$ studied.

For a typical transmission grid as the Scandinavian one, the engineering
textbooks give values of $\alpha\approx 0.24$ (see Methods section) as realistic.
For both network models, even much smaller values lead to a
dramatic change in the basin structure. Increasing $\alpha = 0$ to $0.05$, the
basin of solitary states disappears, only to reappear at $\alpha = 0.15$ but
mirrored as exotic solitaries, rotating contrary to the natural limit cycle.
For higher values of $\alpha$, other states start playing a significant role,
and the behaviour of the circle model and the Scandinavian topology start to
differ.

\begin{figure*}[!ht]
  \begin{subfigure}[b]{0.26\textwidth}
    \caption{ASBS - Scandinavian}
    \includegraphics[width=\textwidth]{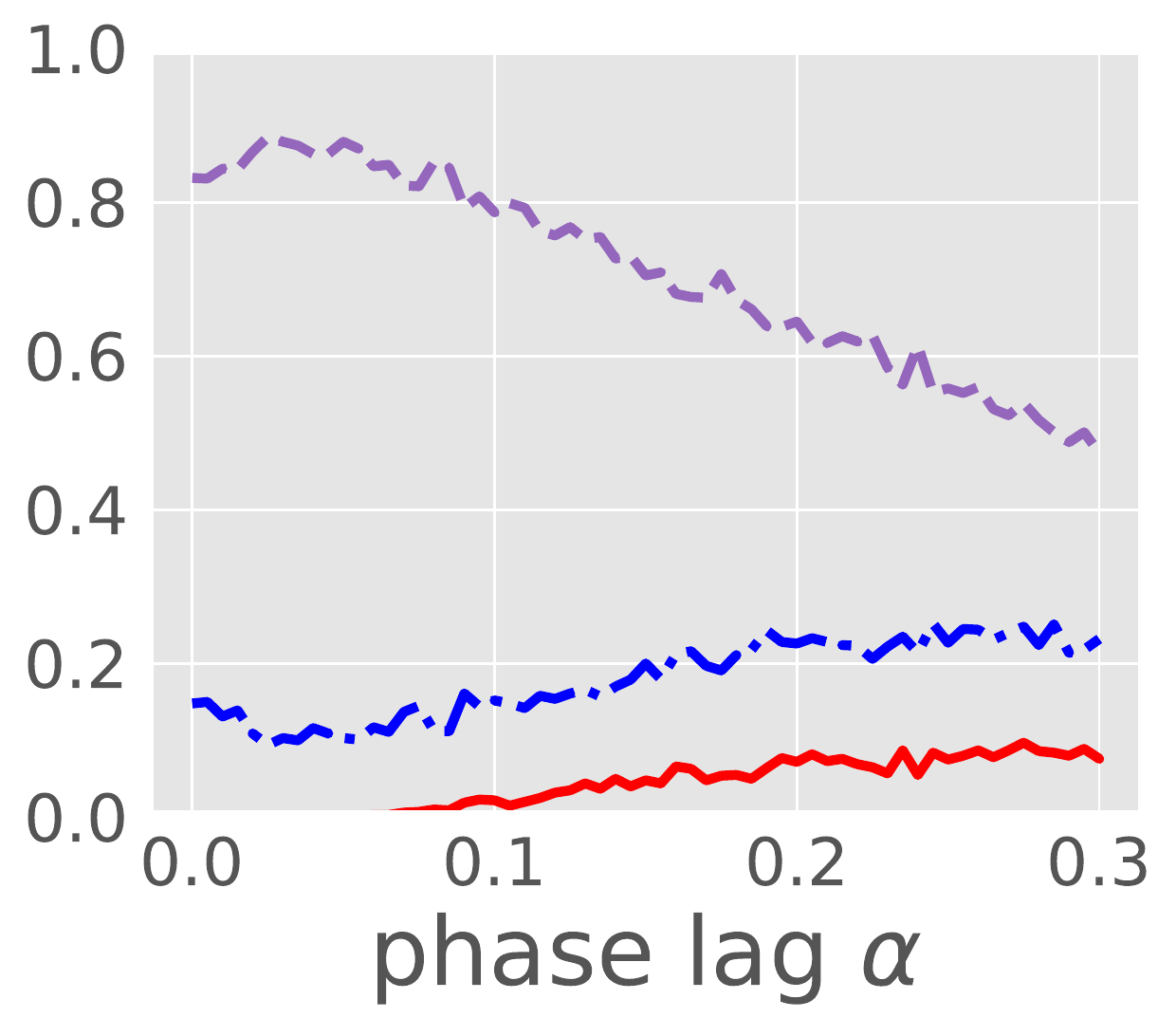}
    \label{fig3a}
  \end{subfigure}
  \begin{subfigure}[b]{0.26\textwidth}
    \caption{ASBS - Circle}
    \includegraphics[width=\textwidth]{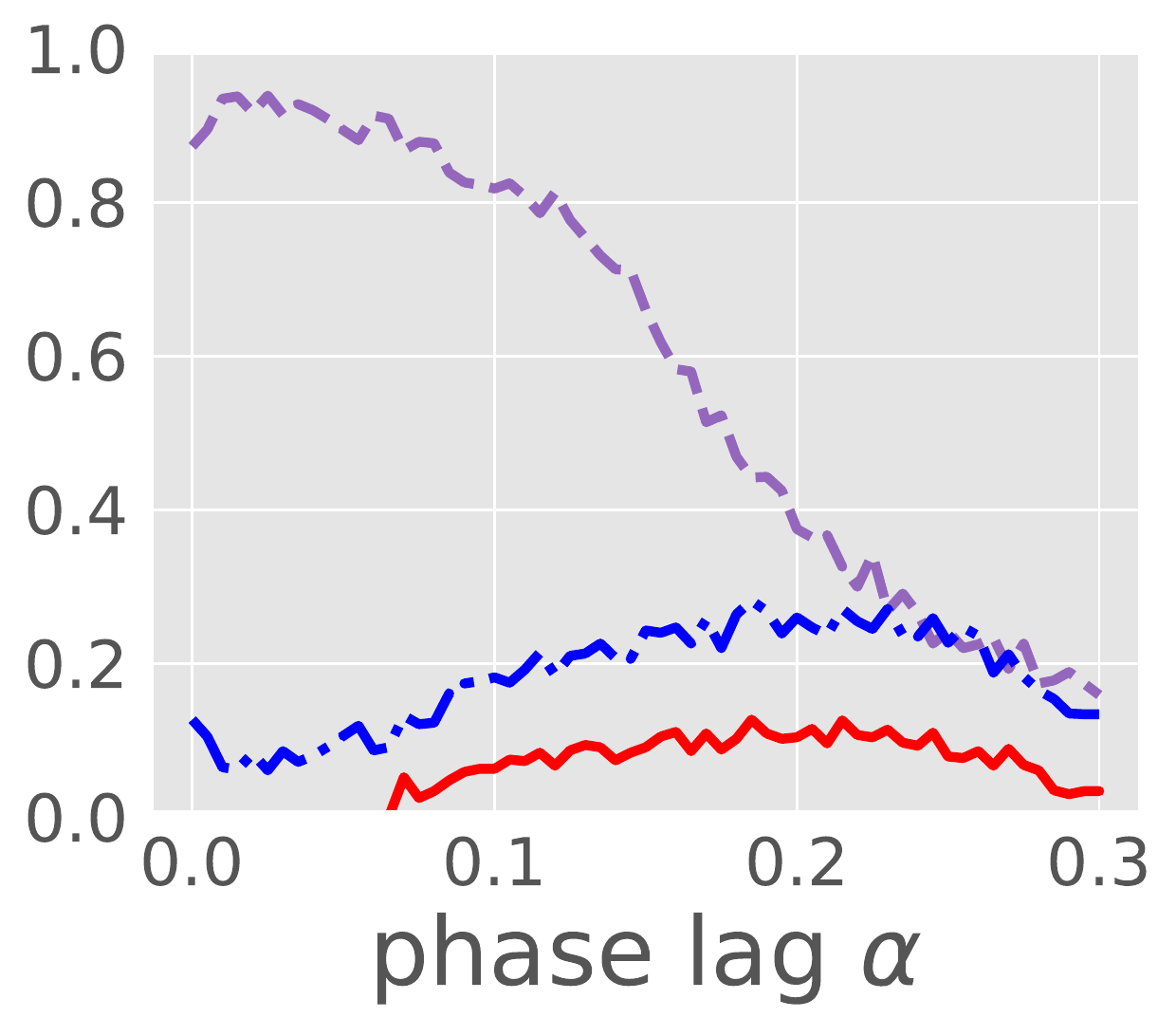}
    \label{fig3b}
  \end{subfigure}   
  \begin{subfigure}[b]{0.26\textwidth}
    \caption{global BS - Circle}
    \includegraphics[width=\textwidth]{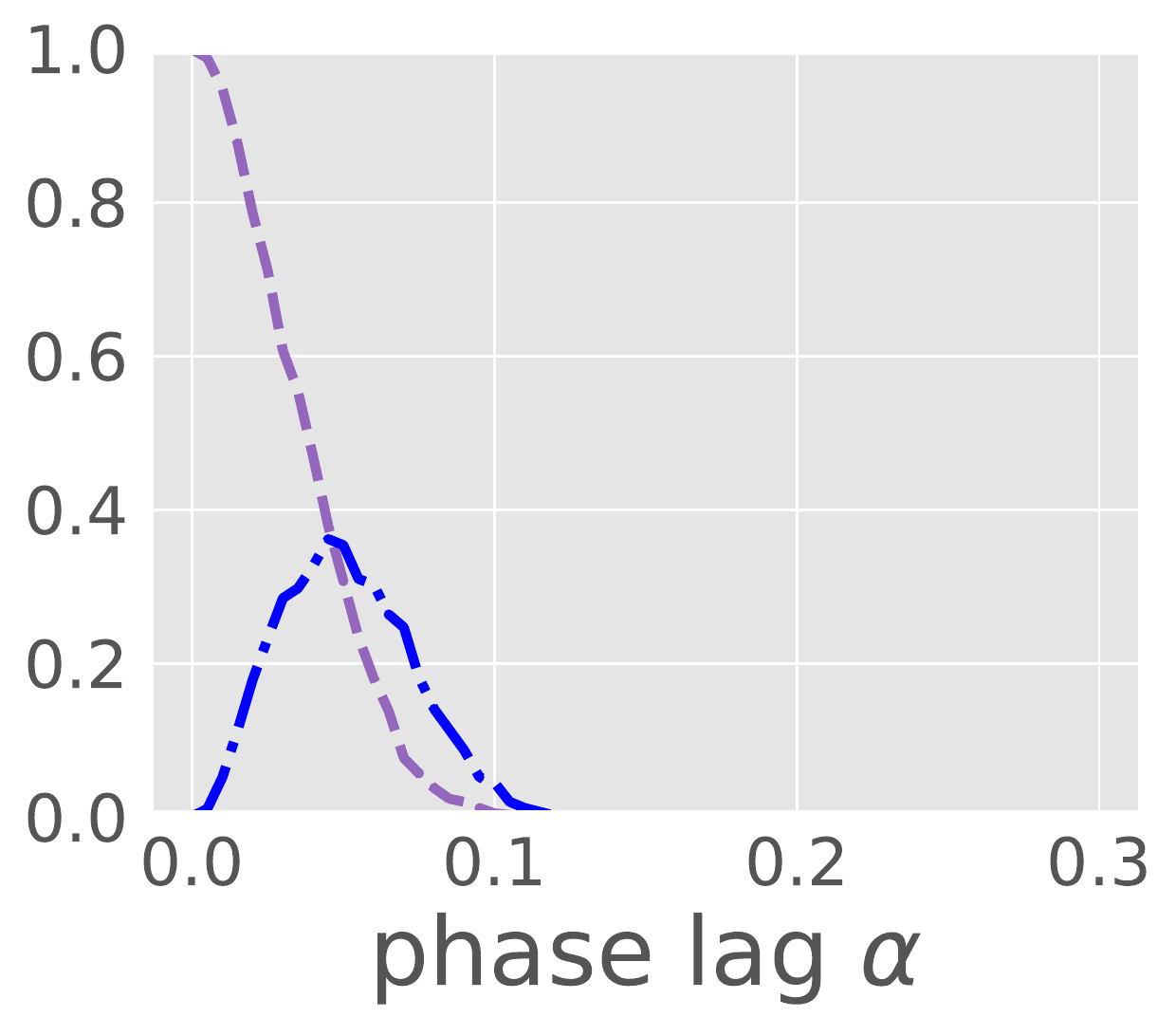}
    \label{fig3c}
  \end{subfigure}

  \begin{subfigure}[b]{0.26\textwidth}
    \caption{high coupling}
    \includegraphics[width=\textwidth]{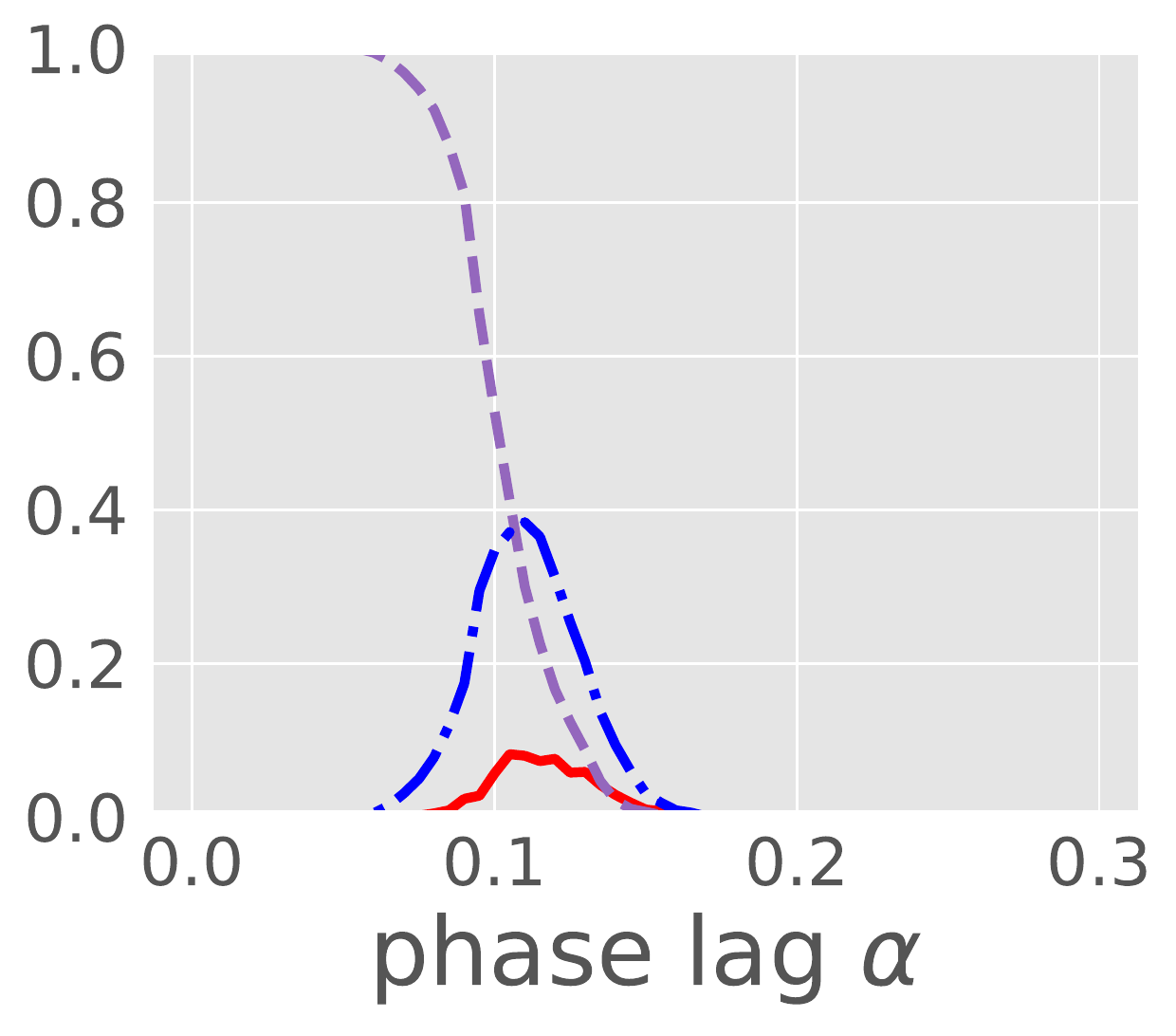}
    \label{fig3d}
  \end{subfigure}
  \begin{subfigure}[b]{0.26\textwidth}
    \caption{medium coupling}
    \includegraphics[width=\textwidth]{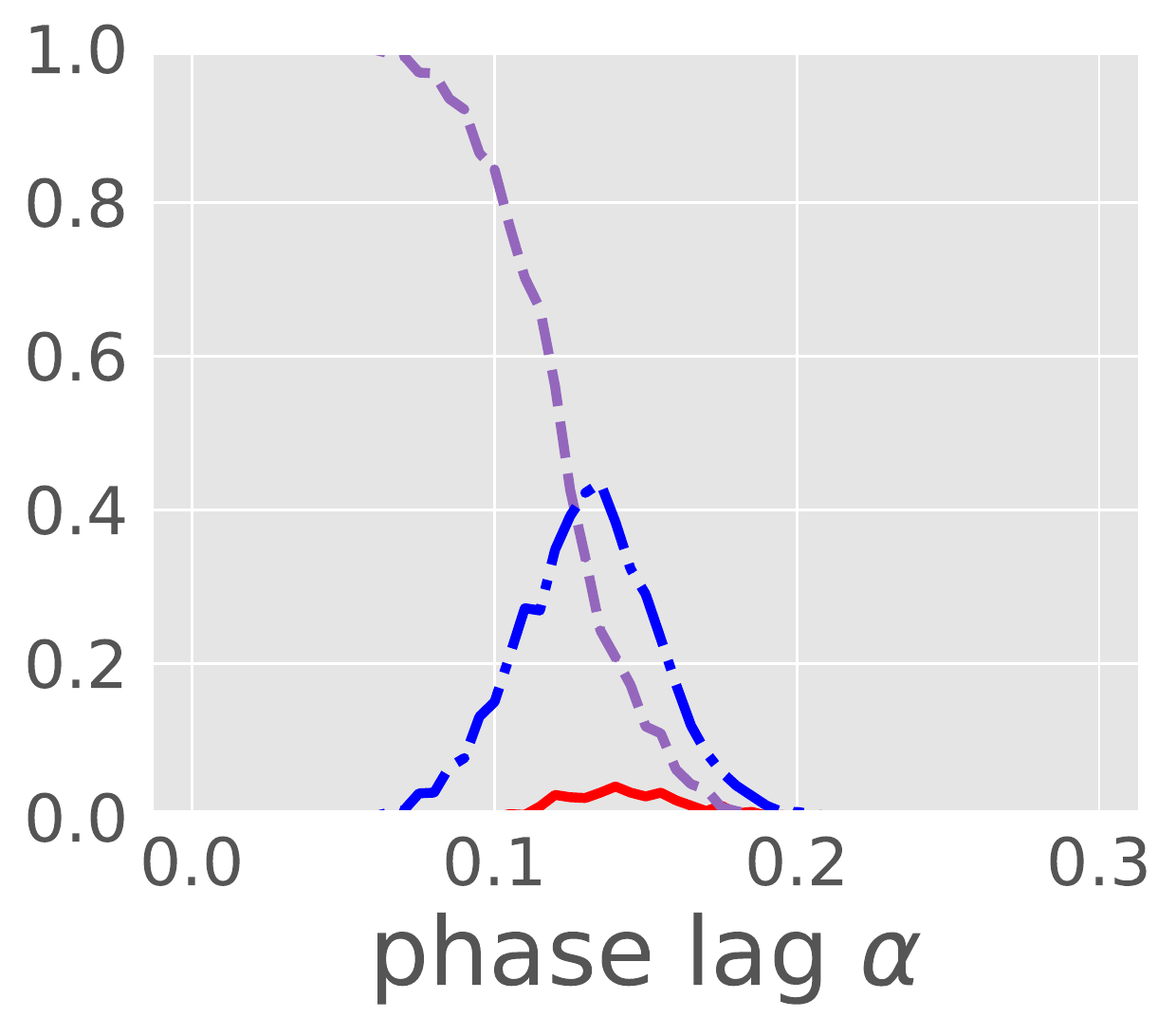}
    \label{fig3e}
  \end{subfigure}
  \begin{subfigure}[b]{0.26\textwidth}
    \caption{low coupling}
    \includegraphics[width=\textwidth]{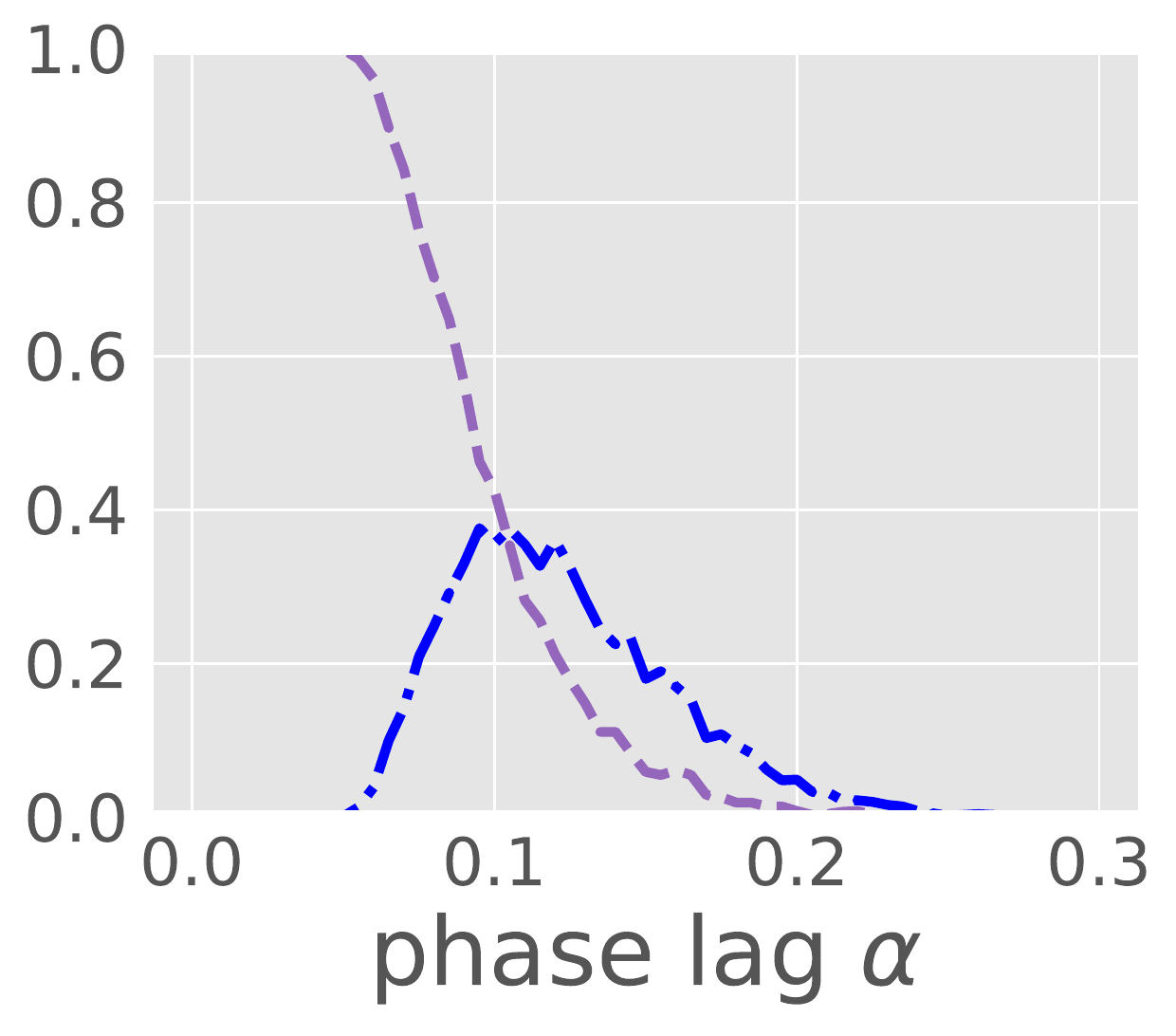}
    \label{fig3f}
  \end{subfigure}

  \begin{subfigure}[b]{0.26\textwidth}
    \includegraphics[width=\textwidth]{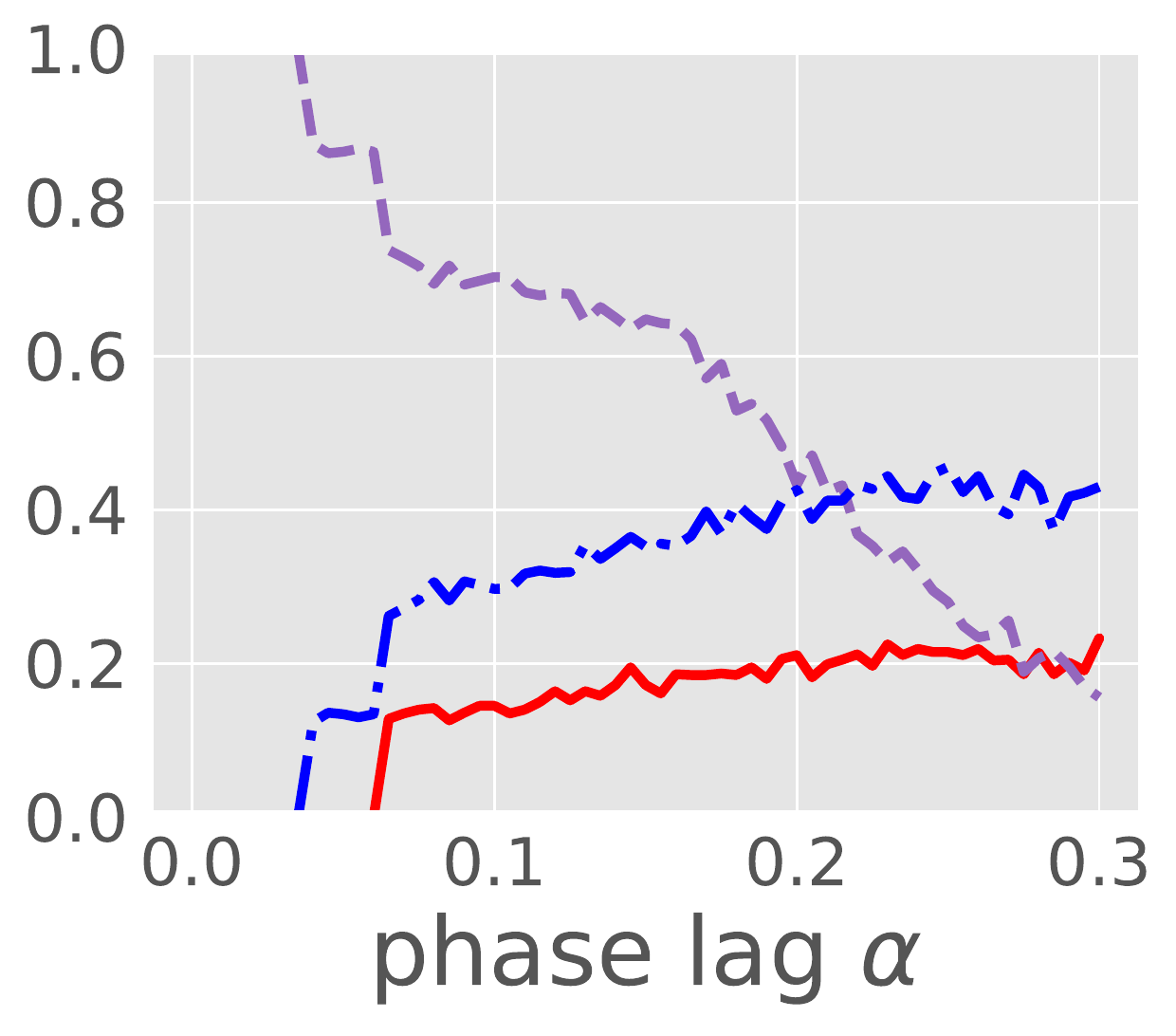}
  \end{subfigure}
  \begin{subfigure}[b]{0.26\textwidth}
    \includegraphics[width=\textwidth]{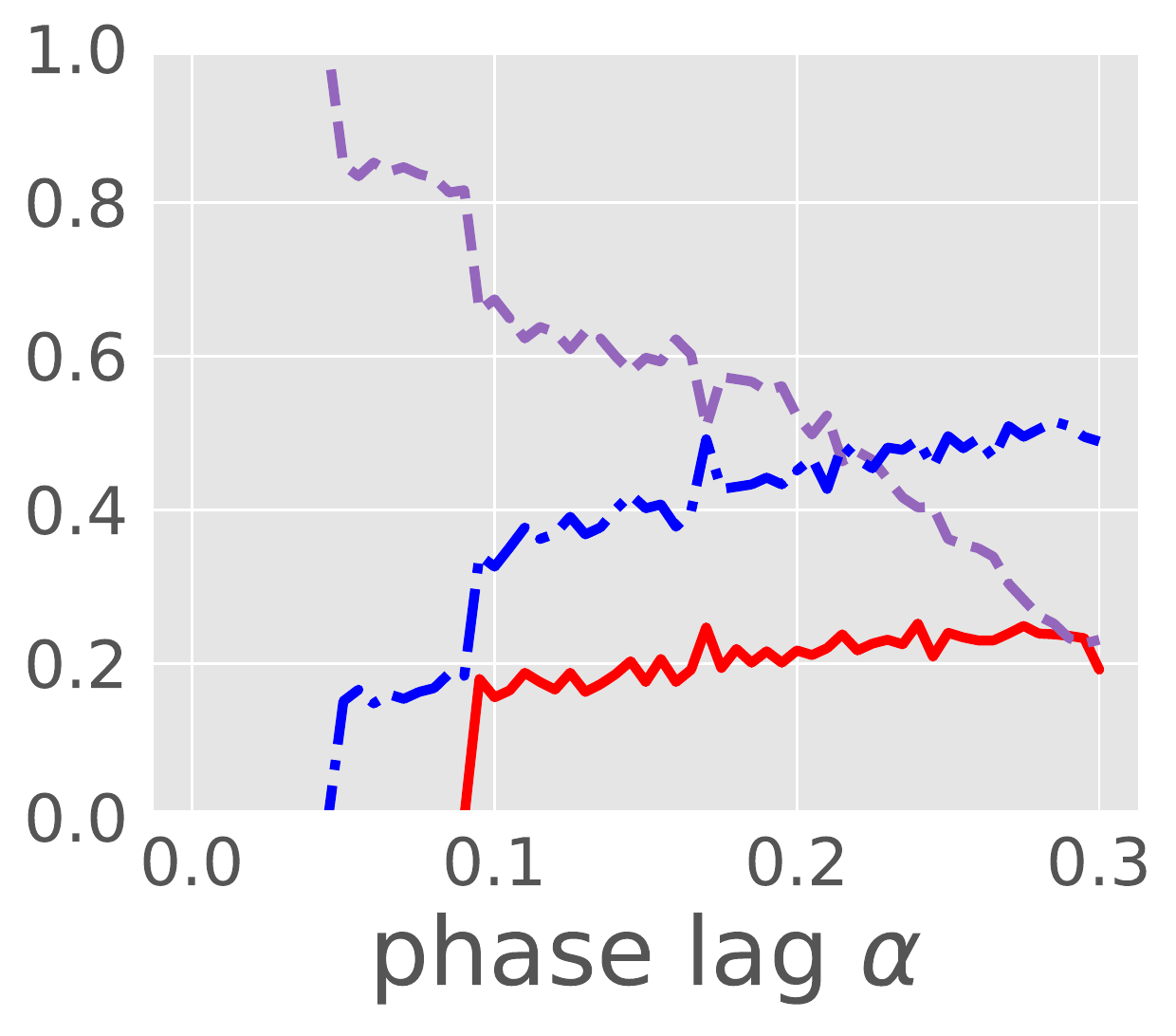}
  \end{subfigure}
  \begin{subfigure}[b]{0.26\textwidth}
    \includegraphics[width=\textwidth]{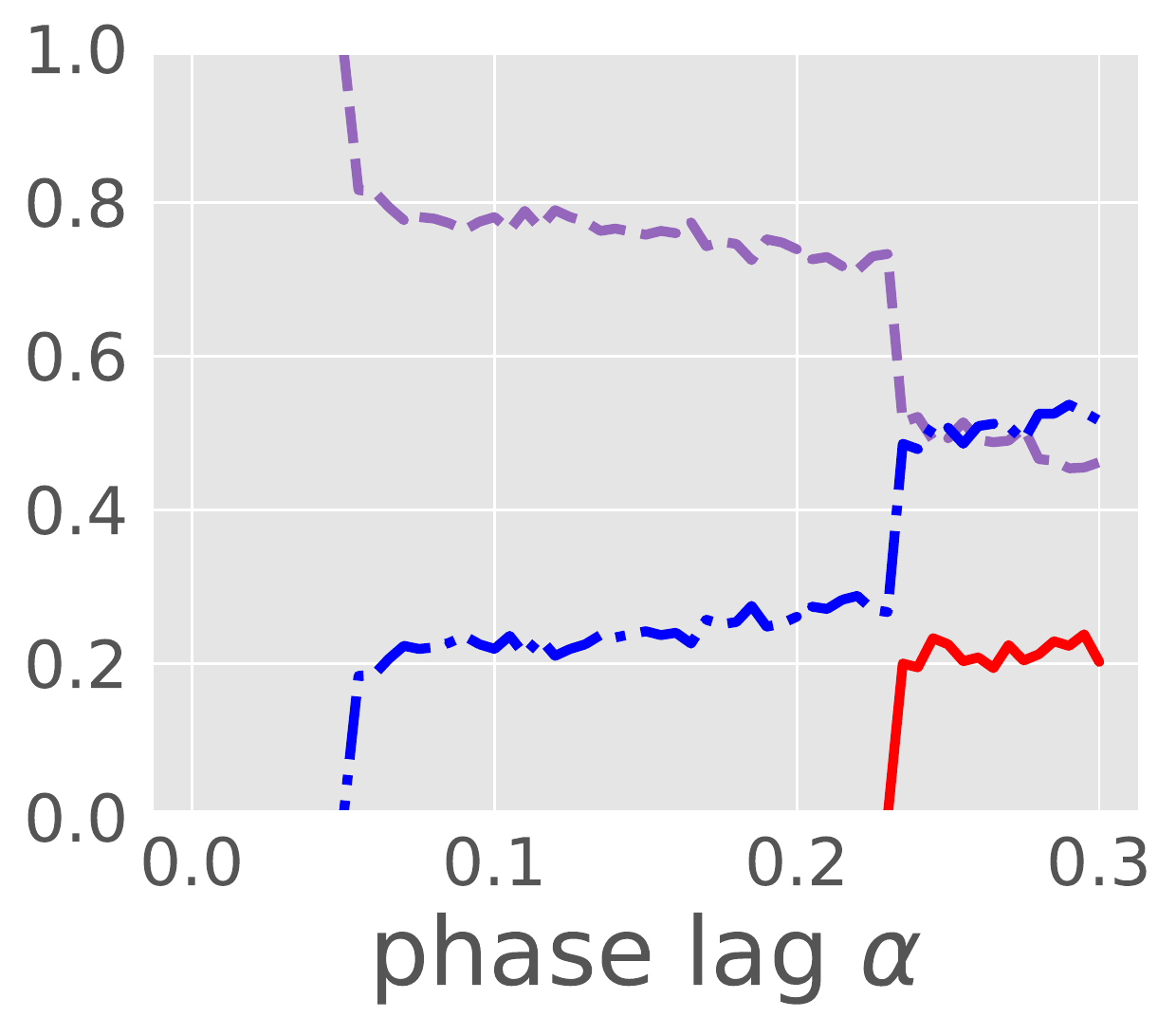}
  \end{subfigure}
  \caption{\textbf{Basin stability.}
  The top row shows the average single node basin stability ASBS (\textbf{a},
  \textbf{b}) and the global basin stability BS (\textbf{c}) of three types of
  asymptotic regimes: synchronisation (\dashed), exotic solitaries
  (\full) and the union of normal and exotic solitaries (\chain). 
  Simulations were performed with standard parametrisation and control (see Methods).
  \textbf{d}-\textbf{f} global BS (top) and ASBS (bottom) for the circle
  topology with enhanced control and different different coupling strengths. 
  Line styles as in \textbf{a}-\textbf{c}.
  Further parametrisations are given in SI.
  }
 \label{fig3}
\end{figure*}

To quantitatively study this effect, we consider the average single node basin
stability (ASBS) for these systems,  i.e., the probability that a perturbation
at a random node will lead back to the synchronous, to a solitary, or an
exotic solitary state. The ASBS of the synchronous state is a proxy for
realistic large perturbations that will typically be geographically localized
in the power grid, rather than perturb the system globally. We also study
smaller global perturbations and likewise define a global basin stability. For
a more detailed discussion of the probabilistic stability measures used here
see the Methods section. Figs.~\ref{fig3a} and \ref{fig3b} show that the ASBS
is decreasing dramatically for both models, the Scandinavian and the circle
topology, confirming the observation from the single node phase space pictures
of the top two rows. In the Scandinavian power grid we see a reduction of ASBS
from  values larger than $0.8$ for $\alpha = 0$ all the way down to $0.5$ at
$\alpha = 0.3$. At the same time the ASBS of solitary states is somewhat
enhanced, rising from a low of $0.1$ to above $0.2$. The circle topology
exhibits an even more pronounced collapse of ASBS and enhancement of the basin
of solitaries, with ASBS values as low as $0.2$ for realistic values of
$\alpha$.  The global basin stability of the synchronous state is numerically zero in
the Scandinavian topology and not shown, but for the circle topology we see a
dramatic and complete collapse of stability as $\alpha$ is switched on. The
stability of the synchronous state is reduced from $1$ to $0$ for
unrealistically small $\alpha$ of $0.1$ already. In the intermediate regime
the basin of solitaries takes over, at higher $\alpha$ we typically have
several solitaries or larger clusters falling out of synchrony (see SI).

To systematically study the basin and existence of the synchronous and
solitary state, we will analyse the simpler, highly symmetric circle topology
further. As above, for realistic parameter choices the results are
qualitatively the same for this choice of topology as for the Scandinavian
power grid. In order to explore other dynamical regimes we also investigate
the circle with a parametrization where the damping is much enhanced compared
to the coupling and power (see Method section for details). this corresponds
to an enhanced reaction of the droop control that adjusts power as a reaction
to a frequency deviation. We further investigated different ratios of injected
power to coupling strength, leading to differently loaded lines. In all
dynamical regimes the circle topology also allows us to study the global basin
stability. The detailed results are contained in the SI.

In Fig.~\ref{fig3} we find that the global basin stability starts at $1$ for
all couplings when we have enhanced control and collapses completely when
$\alpha$ is increased to realistic values. We also see that the collapse
proceeds via the occurrence of a global basin of solitary states. Whereas for
ASBS it is somewhat natural to expect solitary states to play a prominent
role, given that only one oscillator is perturbed, here we uncover that the
solitary states play a prominent role in reducing the size of the sync basin
\cite{Wiley2006}  in the full phase space as well.

The ASBS plots in Figs.~\ref{fig3d}, \ref{fig3e} and \ref{fig3f} now can be
interpreted by considering the shifting of the limit cycle
$\omega_{LC}^\prime$ with $\alpha$. While the limit cycle is not stable at
$\alpha = 0$, as we increase $\alpha$ its frequency gets shifted to
increasingly negative values. For oscillators that are naturally rotating
negatively, the limit cycle gets shifted into the parameter regime where it
becomes stable first. At higher $\alpha$ the oscillators that are naturally
rotating positively also gain a stable negatively rotating limit cycle, i.e.
the exotic solitary described above. Note that the onset of exotic solitaries
occurs at decreasing $\alpha$, i.e. for smaller losses, although the  coupling
strength is increased. The occurrence of other solitaries, however, is less
sensitive to the coupling.

\begin{figure}[!ht]
   \begin{subfigure}[b]{0.48\columnwidth}{}
    \caption{Consumer, standard parametrization}
    \includegraphics[width=\textwidth]{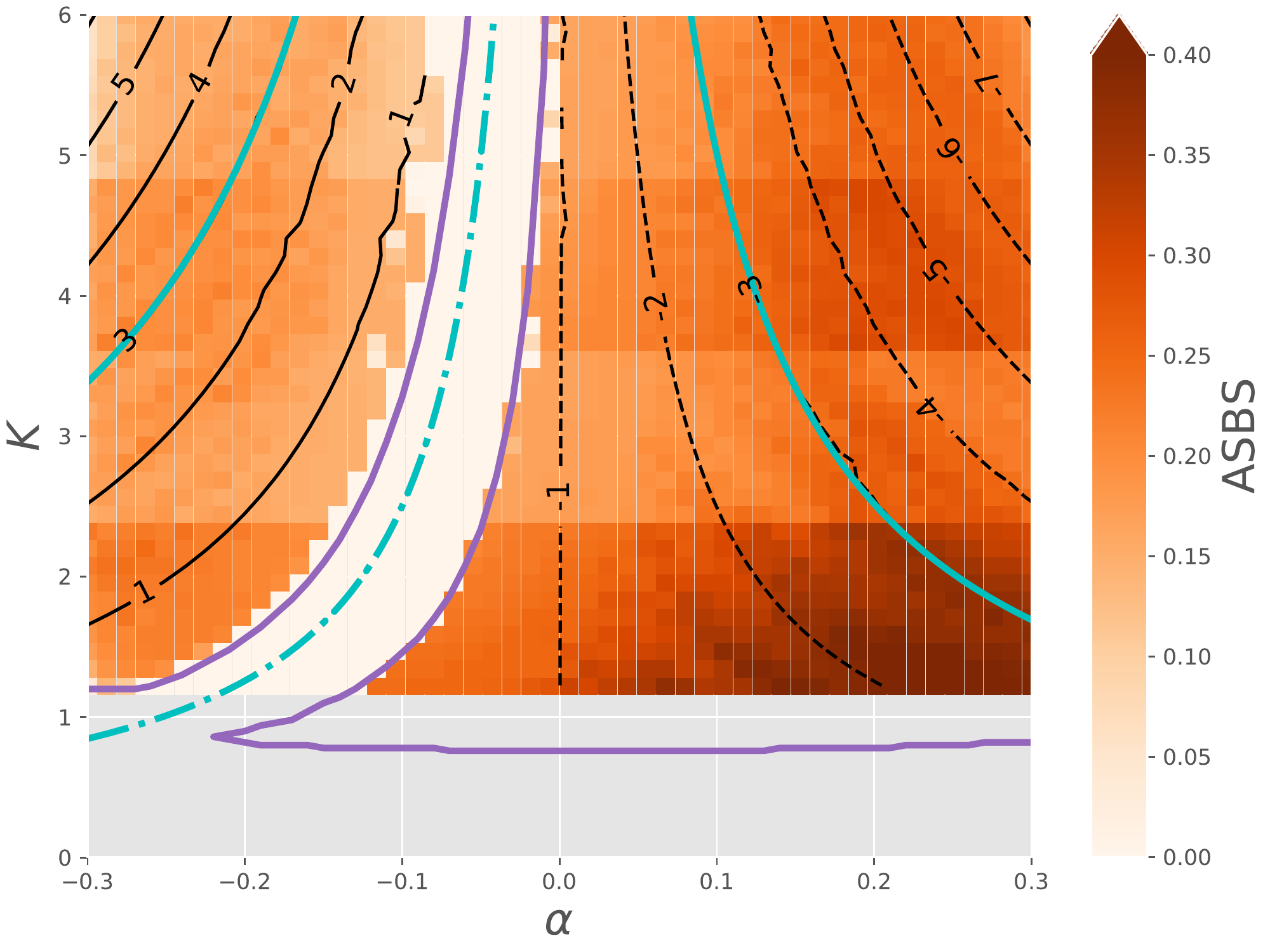}
    \label{fig4a}
  \end{subfigure}
   \begin{subfigure}[b]{0.48\columnwidth}
    \caption{Producer, enhanced control (P=0.1)}
    \includegraphics[width=\textwidth]{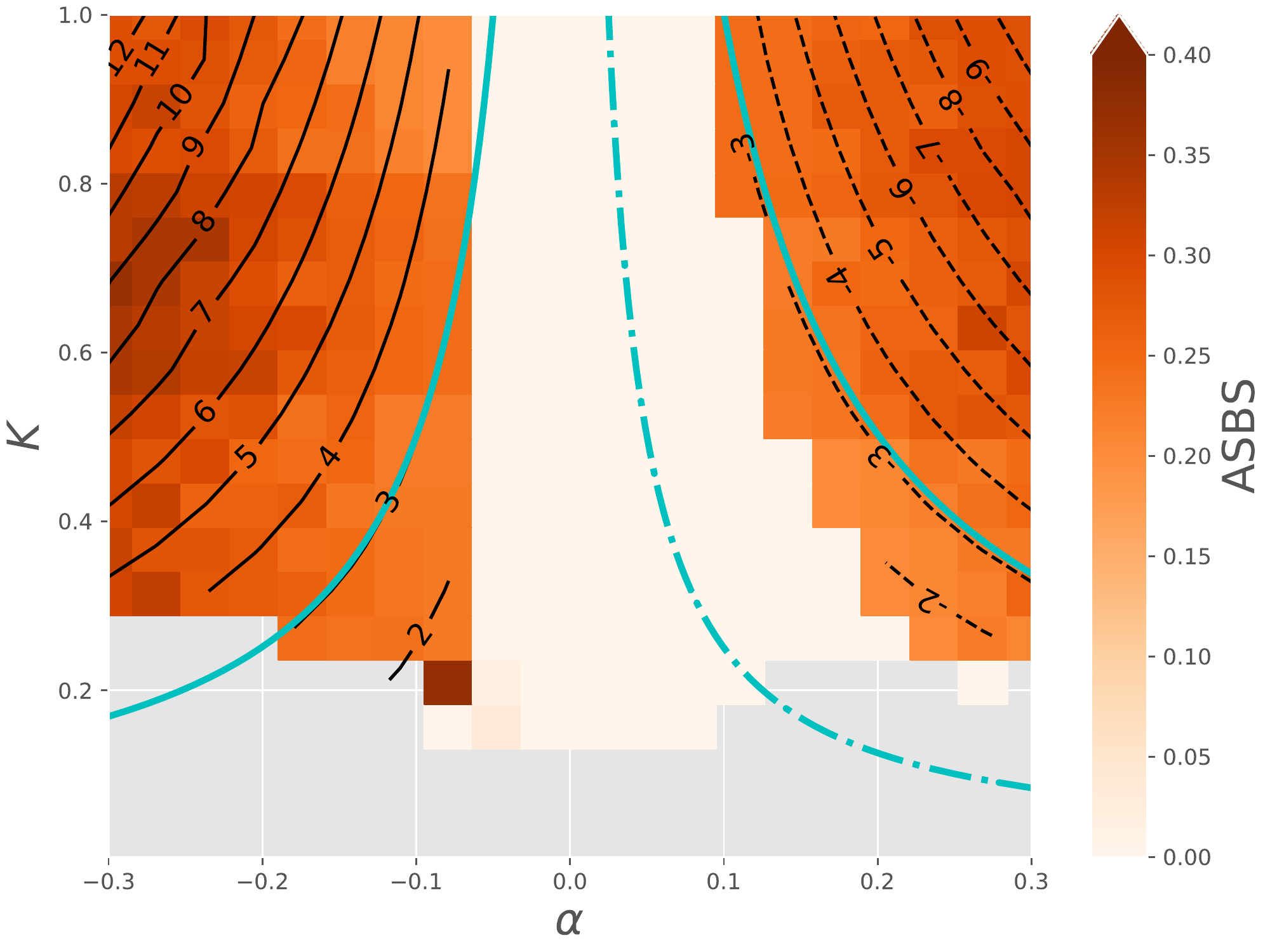}
    \label{fig4b}
  \end{subfigure}
  \caption{\textbf{Existence regions.}
  $K$-$\alpha$ parameter plane for solitary oscillations in the circle
topology   at \textbf{a} a   consumer (standard parametrization) and
\textbf{b} a producer    (enhanced control). The colour scale indicates the
average single node basin stability ASBS    of normal respectively exotic
solitaries.    The contour lines give the mean instantaneous frequency of the
observed solitaries as an integer multiple of the natural limit cycle
$\omega_\text{LC}$, positive for solid, negative for dashed line style. The
cyan solid lines (\inf) show the analytic location of the
$\vert\omega_\text{LC}^\prime\vert=3\vert\omega_\text{LC}\vert$ limit cycle
from the infinite bus model. As a special case, the dash-dotted cyan line
(\infzero) marks the respective hypothetical location of the
$\omega_\text{LC}^\prime=0$ limit cycle.  \textbf{a} Additionally, the purple
lines (\cont) mark the extend of the solitary existence regions determined by
numerical state continuation (see Methods).
Complementary plots for producers/consumers are given in SI.
}
 \label{fig4}
\end{figure}

In order to confirm this interpretation we show in Fig.~\ref{fig4} a
systematic study of the ASBS in the $K$-$\alpha$ parameter plane, both for the
standard parametrization and for the enhanced control.  Note that data points
below the line $K\approx P$ are excluded, since ASBS is not defined
in the absence  of the synchronous fixed point (see Method section for details
of the numerical procedure). For comparison, we also show various markers of
the infinite bus model (Eqn.~\ref{eq:infinite-bus-model}) with shifted limit
cycle $\omega_{LC}^\prime$.  We find that the frequency of the solitary is
indeed explained by the shifted frequency of the infinite bus model.
Additionally, the disappearance of the solitary and the later appearance of
the exotic solitary state can be understood as $\omega_{LC}^\prime$ being
shifted close to zero, and then beyond. This suggests that both transitions
from synchronisation to solitary oscillations are homoclinic bifurcations,
which occur for the infinite bus model. Numerically, this is
supported by a logarithmic scaling  of the solitary's oscillation period and
the mean-field approximation (see SI and the discussion in
\cite{jaros2018solitary}).

Thus for all coupling strengths in the realistic regime, there is a strong
qualitative difference in the asymptotic structure between the lossless case
$\alpha = 0$ and realistic values for $\alpha$. Further, while the bimodal state, where
all nodes oscillate close to their natural frequency $\omega_{lc}$, exists
only for moderately small coupling (see SI, Fig.~\ref{fig:asbs-further}),
solitary states persist even for coupling strengths much beyond realistic
values. In fact, the numerical continuation of solitary states in the
parameter plane suggests that, while the existence region for exotic
solitaries is  bounded by a maximum coupling strength, this is not
the case for normal solitaries.
Due to the shape of the region of existence of solitary states,
an increase in coupling can even lead to exotic solitaries coming into existence.

Finally, note that, next to the single solitary states studied
in the bifurcation diagram of Fig.~\ref{fig4}, many other asymptotic states,
including ones with larger desynchronized clusters and several disjoint
solitaries exist. These start to play a role for larger $\alpha$ as in Fig.~\ref{fig2b} (see also \cite{jaros2018solitary}, SI).

In summary, for all studied parameter regimes -- including the case of a realistically
parametrized Scandinavian system -- we observe a dramatic reduction of the size
of the basin of attraction of the synchronous desirable operating state when going to realistic values of losses. To
our knowledge, this very generic phenomenon has not been observed in the
literature before. It has far reaching implications for the design and control
of future power systems. Our results show that it is not possible to use
theoretical results on the stability of systems without losses ($\alpha = 0$) in the
design of future power systems as is. The losses fundamentally alter the
dynamical behaviour of the coupled systems.

While we find that an enhanced control effort is not sufficient to suppress solitary states,
a strong control regime, with an
extremely high ratio $\frac{D^2}{P H}$ of a hundred times that of the
standard value (see Methods), can stabilize the system up until $\alpha\lessapprox 0.2$ (see SI,
Fig.~\ref{fig:week1/northern hd 0}). Beyond this, however, for realistic values of $\alpha\approx 0.24$ the global stability is already rapidly collapsing while ASBS is slowly decreasing. For future distribution grids with $\alpha$ much larger than $0.24$ even such a probably unrealistic, and for other reasons undesirable,
control effort is not sufficient to eliminate the multistability induced by the losses.

Further, as noted above, a mere
increase of coupling, e.g. by improving transmission capacities in a power
grid, does not suffice to prohibit non-synchronous solitary oscillations either.
In some specific situations it can even enhance their basin.
Controling such lossy dynamical networks and solitaries in them thus poses a fundamental challenge,
to future research in network control that can not be answered by tweaks to system parameters.

The study of lossy coupling also has implications
for the engineering literature. While case studies typically work with the physically
correct equations involving losses (e.g. \cite{liu2017quantifying}) it is also typical
to neglect the losses of e.g. coupled two-area systems \cite{ulbig2014impact}.
It remains unclear whether this is justified. Novel control strategies based on basin structure of the lossless case, 
like those in \cite{Vu2017}, will certainly need to be revisited. The degree to which the
presence of losses also affects the transient and stochastic behaviour of the
power grid is also a subject of ongoing research \cite{auer2017stability}.

The fact that this effect appears to be not strongly affected by the network
topology means that further investigations -- including experiments and analytics -- can
begin on simplified systems, and the effect can be cleanly isolated from
pronounced local topological effects present in real world topologies.
In particular, solitary states persist even in the mean-field limit 
with an all-to-all coupling, as we extensively discuss in the SI.

Finally, as noted before, the model studied here can be equivalently seen as representing generic
Newtonian oscillators, coupled by the lowest Fourier mode. When approximating
general systems as phase oscillators (see e.g.
\cite{Nakao2016}), the parameter $\alpha$ appears generically. We
therefore expect the phenomenon described here to pertain to a wide range of
non-linear heterogeneous oscillator systems.

\section*{Methods}

\subsection*{Parametrization of networks and dynamical regimes.}

Typical values for ohmic losses and inductivity in power grids are\cite{auer2016can}:

\begin{center}
  \begin{tabular}{ l | c | c | c }
    \hline
    Line & Transmission & Distribution & Low voltage \\ \hline
    R $[\Omega/km]$ & 0.1 & 0.4 & 0.5 \\ \hline
    X $[\Omega/km]$ & 0.4 & 0.3 & 0.08 \\
    \hline
    $\alpha$ & 0.24 & 0.93 & 1.41 \\
    \hline
  \end{tabular}
\end{center}

Thus the regime studied in this work covers transmission lines, where the lossless approximation has been taken to be valid.

The power grid is then parametrized in accordance with the values most studied
in the basin stability literature
\cite{menck2014dead,nitzbon2017deciphering,schultz2014detours}, $H = 1$, $P =1$, 
$D = 0.1$, and $K \approx 6$ for the geographic power grid, and $K = 6$
for the circle topology.

This parametrisation is typically used in the theoretical physics literature in order to study network and multistability effects expected to be relevant in future power grids. It is not expected to be a very close representation of current power grids, but to reveal structural features of the type of dynamics expected in future power grids. As the observed phenomenon is highly generic, and depends primarily on the physics of the coupling, we expect it to occur in all parametrizations of the swing equation though.

In order to be self-contained we will briefly review the detailed motivation behind the parametrisation used. Currently each node in a transmission grid connects large-scale conventional power plants or serves as the connection point of underlying distribution grids. In the future, increasingly decentral power generation in active distribution grids, means that every transmission node
effectively represents a region with a net power output that can be either positive or negative. Assuming that the location and production of decentral generation is not correlated with the existing grid structure, it is reasonable to simply choose at random whether a region is producing or consuming.

In the classical case, consumer regions are modelled as constant demands and the dynamics occurs purely at the generators. As a reduction of conventional generation reduces system inertia, we expect grid-forming inverter-interfaced distributed generation units to contribute to overall system inertia as well as droop control. Thus load nodes are reasonably modelled by inertial, damped nodes with a net power demand.

We believe that a random assignment of net generation/consumption without pre-assuming the actual amount is a plausible first-order model for the described scenario.

\subsection*{Dynamical regimes.}

The bifurcation structure of the system is time parametrization invariant, and
thus there are two characteristic quantities in the system. Meaningful choices
for time invariant parameter combinations are $\frac{P}{K}$, which
characterizes the load on the lines, and $\frac{D^2}{P H}$ which describes the
strength of the droop control relative to the power infeed.

Next to the standard parametrization, we also studied a strongly damped regime
with enhanced droop control driving the system to the fixed point, and an
extremely strong droop control, for which the results are shown in the SI,
Fig.~\ref{fig:week1/northern hd 0}.

\begin{center}
  \begin{tabular}{ l | c  | c | c }
    & Standard & Enhanced control & Strong control \\ \hline
    \hline
    $\frac{D^2}{P H}$ & $1/100$ & $1/10$ & $1$ \\ \hline
  \end{tabular}
\end{center}

We further varied the coupling strength, and hence the line loading.

\begin{center}
  \begin{tabular}{ l | c | c | c | c }
    & Standard & High & Medium & Low coupling \\\hline
    \hline
    $\frac{P}{K}$ & $\approx 1/6$ & $1/20$ & $1/10$ & $1/2.5$ \\
    \hline
  \end{tabular}
\end{center}

\subsection*{Probabilistic stability anlysis.}

The main numerical method {we apply here is the} 
sampling-based evaluation of probabilistic properties of dynamical systems
\cite{hellmann2016survivability,lindner2018stochastic,menck2014dead,menck2013basin,nitzbon2017deciphering,schultz2018bounding,schultz2017potentials}. 
{In particular, basin stability is the probability that a system converges to a 
specific attractor given a distribution of random perturbations. 
This can be regarded as a volume measure of the attraction basin. 
In the context of synchronisation it became popular as the 'size of the sync basin' \cite{Wiley2006}. 
A direct evaluation of the basin geometry is computationally not feasible in 
high-dimensional systems.
Even if the set is known to be convex, current algorithms scale with $\mathcal{O}(n^4)$ 
where $n$ is the phase space dimension \cite{Lovasz2006a}.
Instead, a Monte Carlo sampling yields efficient 
estimates of the volume, whose standard error does 
not depend on $n$ \cite{menck2013basin}.}

{Compared with mathematical methods based on Lyapunov functions, the advantage and draw back of sampling based methods is that they are not sensitive to worst case scenarios. Instead, they provide accurate results on the typical behaviour of a dynamical system, even where analytic results are hard to achieve or not available.}

The perturbations are drawn uniformly at random from a finite box centered around 
the synchronous state. The synchronous state itself changes with $\alpha$ and 
thus the box of initial conditions does, too. Fig.~\ref{fig5} in the SI
illustrates how the synchronous state changes with alpha.

We use both \textit{global perturbations}, where the perturbations apply to the all
phase space variables, and \textit{network-local perturbations} where
perturbations are localized at a particular node. Unless otherwise specified,
global basin stability was obtained by sampling from initial conditions with
$\Delta \dot \phi_i \in [-0.1 \omega_{lc}, +0.1 \omega_{lc}]$ and  $\Delta
\phi_i \in [-\pi, +\pi]$. Our second measure is the average single node basin
stability, the probability to run to a particular state after a perturbation
at a randomly drawn node, drawn from $\Delta \dot \phi_i \in [-1.5
\omega_{lc}, +1.5 \omega_{lc}]$ and  $\Delta  \phi_i \in [\phi_i^\ast-\pi, \phi_i^\ast+\pi]$.
Here, the $\phi_i^\ast$ denote the phase values at the synchronous fixed point,
i.e. ASBS is determined by perturbing a randomly chosen node's degrees of freedom
away from the fixed point. This particular choice of perturbations is the 
key difference to the determination of global basin stability. Consequently,
ASBS cannot be determined in the absence of a synchronous regime for certain
parameter values.

For the standard configuration the global basin stability is numerically not
distinguishable from zero. One way to interprete this is that each oscillator
has a finite probability to desynchronize and perturbing all at the same time
makes the probability that none of them do exponentially small. The average
single node basin stability is a more robust and more realistic measure of the
stability of the synchronous state.

\subsection*{Continuation Study}

To perform a numerical continuation of solitary states in the parameter plane
(see Fig.~\ref{fig4a}), we integrated the system with a Cash–Karp method
(error $e=10^{-4}$) for $T=5\cdot 10^3s$ under the influence
of small noise (amplitude $d=10^{-5}$) to ensure transverse stability. $T$ is
chosen large enough to exclude transient solitary desynchronisation. From a
starting point, one of the parameters is varied in steps of $\Delta\alpha=0.01$
respectively $\Delta K=0.1$ until the solitary oscillation ceased to exist. Then,
the last step is reversed and the algorithm adds  one step to the other
parameter, tracking a bifurcation line.

\subsection*{Software}

Simulations were performed using the Scipy package \cite{scipy} and the
pyBAOBAP package available at \url{https://gitlab.pik-potsdam.de/hellmann/pyBAOBAP}. Scipy uses the LSODA solver of the odepack
Library \cite{hindmarsh1983odepack}.

All code and data will be published open source at \url{https://gitlab.pik-potsdam.de/pschultz/solitary_power_grid} or is available upon request.

\begin{acknowledgments}
FH, PS and JK acknowledge the support of BMBF, CoNDyNet, FK. 03SF0472A.
PJ has been supported by the Polish National Science Centre,
PRELUDIUM No. 2016/23/N/ST8/00241 and by the Foundation for
Polish Science (FNP), START Programme, TK has been supported by
the Polish National Science Centre,
MAESTRO No. 2013/08/A/ST8/00/780.
This work was supported by the Volkswagen Foundation  (Grant No. 88462).
Funded by the Deutsche Forschungsgemeinschaft (DFG, German Research
Foundation) – KU 837/39-1 / RA 516/13-1. All authors gratefully acknowledge
the European Regional Development Fund (ERDF), the German Federal Ministry of
Education and Research and the Land Brandenburg for supporting this project by
providing resources on the high performance computer system at the Potsdam
Institute for Climate Impact Research. 
\end{acknowledgments}


%

\appendix

\section*{Supplemental Information (SI)}

\subsection*{Derivation of the Power Flow Equation}

Here, we derive the lossy power flow equation (Eqn.~2). $P_{ij}$ should be the
outgoing flow at the node such that the power flow equation reads $P_i =
\sum_j P_{ij}$. According to Ohm's law, the current on a line is given by
$\frac{V_i - V_j}{R_{ij} + i X_{ij}}$, (if the own voltage is higher, current,
and hence energy flows away) or $Y_{ij}(V_i - V_j)$. As $R_{ij}$ and $X_{ij}$ 
are positive, $Y_{ij}$ has positive real part and negative imaginary part. The
Laplacian $L_{ij} = \delta_{ij} \sum_{k} Y_{ik} - Y_{ij}$ then provides us with
the total outgoing current: $I = L \cdot V$. To connect with the literature
written in terms of phase lags $\alpha$, we set $Y = - i K e^{i \alpha}$. Then
$K e^{i \alpha} = i Y$ has a positive real part and a positive imaginary part.
Hence $K$ and $\alpha$ are positive.

Using $V_i = e^{i\phi_i}$, $V_iV_i^\star=1$, the power flow can then be written as follows:
\begin{align}
\sum_j P_{ij} & = \Re(V_i I_i^\star) \nonumber\\
         & = \Re \left( \sum_j V_i L_{ij}^\star V_j^\star\right) \nonumber\\
         & = \Re \left( V_i V_i^\star \sum_k ( - i e^{- i \alpha_{ik}} K_{ik})^\star\right) - \Re \left( \sum_j V_i (- i e^{- i \alpha_{ij}} K_{ij})^\star V_j^\star\right)\nonumber\\
         & = \Re \left( i V_i V_i^\star \sum_k e^{- i \alpha_{ik}} K_{ik}\right) - \Re \left(i  \sum_j V_i e^{- i \alpha_{ij}} K_{ij}V_j^\star\right)\nonumber\\
         & = - \Im \left(\sum_k K_{ik} e^{- i \alpha_{ik}} \right) + \Im \left(\sum_j K_{ij} e^{i( \phi_i - \phi_j - \alpha_{ij})} \right)\nonumber\\ 
         & = - \sum_k K_{ik} \sin(- \alpha_{ik}) + \sum_j K_{ij} \sin( \phi_i - \phi_j - \alpha_{ij}) \nonumber\\ 
         & = \sum_j K_{ij} \left( - \sin(- \alpha_{ij}) + \sin( \phi_i - \phi_j - \alpha_{ij})\right) \nonumber\\ 
\end{align}

\subsection*{Synchronous Regime with Varying Losses}

\begin{figure}[ht]
   \begin{subfigure}{0.48\columnwidth}
   \caption{Instantaneous frequency}
    \includegraphics[width=\textwidth]{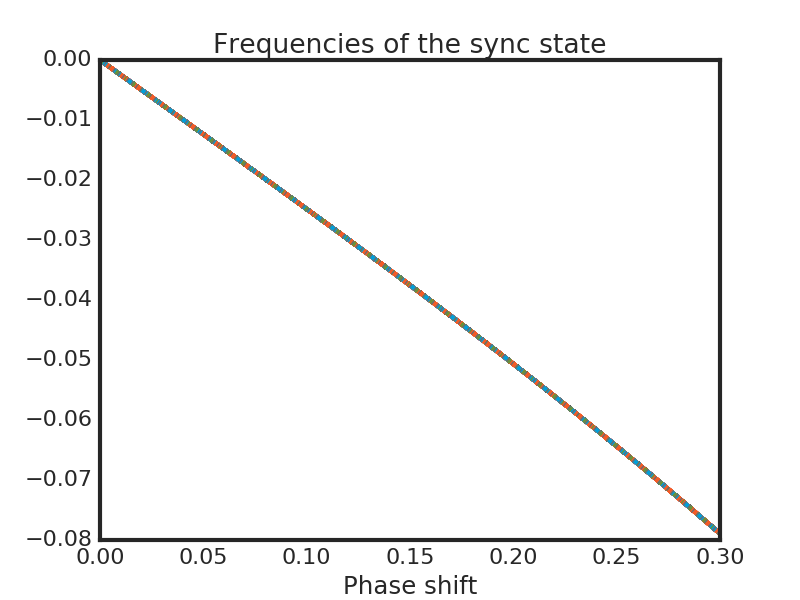}
  \end{subfigure}
   \begin{subfigure}{0.48\columnwidth}
   \caption{Phases}
    \includegraphics[width=\textwidth]{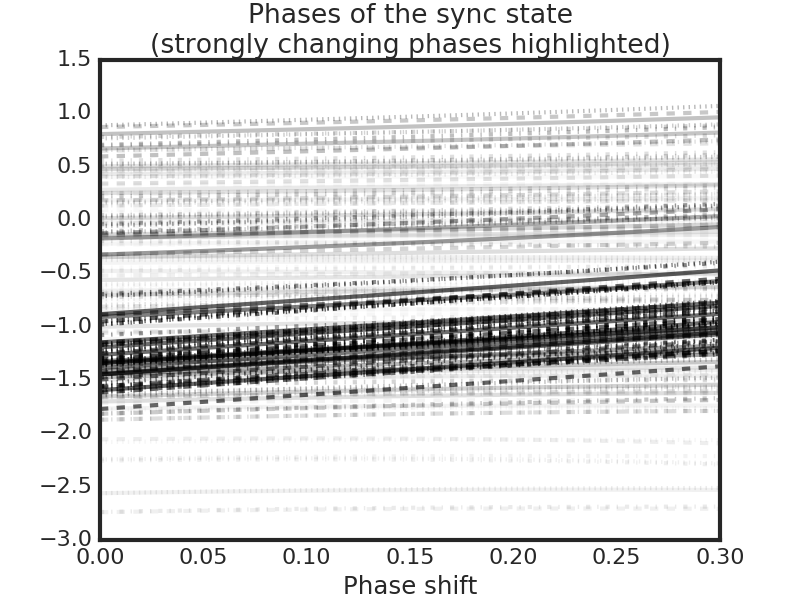}
  \end{subfigure}
  \caption{\textbf{Synchronous regime.} Variation of the synchronous frequency (\textbf{a})
  and phases (\textbf{b}) with increasing phase shift $\alpha$.}
  \label{fig5}
\end{figure}

In Fig.~\ref{fig5} we depict the shift of the synchronous operating point
under an increase of losses which appear as a phase shift $\alpha$ in Eqn.~\ref{eq:power grid}.
The simulation was performed for the Scandinavian power grid with standard parametrization and
enhanced control.

\subsection*{Circle  Model of the Power Grid}

In this section we consider an abstract power grid organization, the so-called {\it
circle model}, which demonstrates the main features we find responsible
for prompt desynchronization. In this model, the desired synchronized state exists
and is stable. However, the model itself appears to be genuinely multistable.
There arise indeed many other, so-called {\it solitary states} in which one or
a few power grid units start to behave differently with respect to all others
perfectly synchronized such that not only the phase but also the average frequency
deviates from the synchronous behaviour. Remarkably, these solitary states are
Lyapunov stable too and hence, they cannot be excluded from the global system
dynamics even by an essential increase of the inner network connectivity (as
it would be naturally expected). Furthermore, it is found that this kind of
network multistability arises under realistic operating parameters applied to
power grids.  If so, there is always a danger that sudden non-small
disturbances associated e.g. with fluctuations in production or demand may
trigger desynchronization of one or several power generators/consumers even if
the synchronous state is perfectly operating in a stable regime. While our
model is topologically simplified and does not account for complication of the
real power grid structure, we show through extensive numerical experiments
that our predictions remain robust when including these effects for the
Scandinavian power grid.

Define the circle power grid as follows.  Let it contains $N$ units and let a half
of them, $N/2$, be identical generators (representing power plants) and the
other half identical motors (representing consumers). Place  all of them
on a ring with the ”alternative” disposition, Fig.~\ref{fig1b}, and assume that each
unit is connected with $R$ of its nearest neighbours to the left and to the right,
and with an equal  coupling strength $K$. In the case, all units obey the same
type of equation of motion, the Kuramoto model with inertia,  of the form

\begin{align}
\ddot \phi_i+ \epsilon \dot \phi_i &= (-1)^{i+1}\Omega + 2RK \sin\alpha +K\sum\limits_{j=i-R, j\neq{i}}^{i+R} \sin(\phi_j - \phi_i + \alpha)\;
\label{sync1}
\end{align}

($i=1,\dots,N$), where the term $K \sin\alpha$ in the right-hand side arises due
to assigned self-coupling topology of the model connectivity. It shifts the
oscillator eigenfrequency $\Omega$ up or down depending on the sign of the
parameter $\alpha$.

Here, we use the notation

\begin{align}
\begin{split}
\Omega &:= \vert P_i\vert/H,\;\;\forall\; i  \\
\epsilon &:= D_i /H,\;\;\forall\; i \\
K &:= K_{ij}/H,\;\;\forall\; i,j \;.
\end{split}
\end{align}

In the lossless case $\alpha=0$ this term disappears and one comes to
the classical Kuramoto model with bi-modal frequency distribution

\begin{align}
\ddot \phi_i+ \epsilon \dot \phi_i &= (-1)^{i+1}\Omega +K\sum\limits_{j=i-R}^{i+R} \sin(\phi_j - \phi_i)\;
\end{align}

which dynamics is comparably simpler. Introducing non-zero $\alpha$, however,
makes the global network behaviour more involved due to multiple appearance of
solitary states, not avoidable even with large increase of the coupling
strength $K$.

\subsection*{Synchronous State of the Circle Model}

To identify a synchronous state of the circle power grid model (Eqn.~\ref{sync1}), re-write it
in a two-group form denoting the generators by $\theta_i$ and the consumers
$\psi_i, ~i=1,\dots,N/2$:

\begin{align}
\ddot \theta_i+ \epsilon \dot \theta_i &= +\Omega + 2RK\sin\alpha +K\sum\limits_{j=i-R, j\neq{i}}^{i+R} \sin(\theta_j - \theta_i + \alpha),\nonumber\\
\ddot \psi_i+ \epsilon \dot \psi_i &= -\Omega + 2RK\sin\alpha +K\sum\limits_{j=i-R, j\neq{i}}^{i+R} \sin(\psi_j - \psi_i + \alpha).
\label{two-group}
\end{align}

Assume $\theta_1=\dots=\theta _{N/2}\equiv\theta$ and $\psi_1=\dots=\psi _{N/2}\equiv\psi$, 
and subtract the second equation from the first one. The following equation is obtained for the phase difference $\eta=\theta-\psi$:

\begin{equation}
\ddot {\eta}+ \epsilon \dot {\eta}=2\Omega-RK(\sin(\eta +\alpha)+\sin(\eta-\alpha)),
\end{equation}

By simple trigonometry, it can be re-written as

\begin{equation}
\ddot {\eta}+ \epsilon \dot {\eta}=2\Omega-\mu\cos\alpha\sin\eta,
\label{Sync2}
\end{equation}

where a new coupling parameter $\mu=2RK$ is introduced. This is an equation of
motion in the two-dimensional invariant manifold indicated by the conditions
$\theta_1=\dots=\theta _{N/2}\equiv\theta$, $\psi_1=\dots=\psi_{N/2}\equiv\psi$ 
as above. Synchronous states of the original $N$-dim
Eqn.~\ref{sync1} are given by the roots of its right-hand side, which are

\begin{equation}
\eta_O=\arcsin\frac{2\Omega}{\mu\cos\alpha}, ~~~~ \eta_S=\pi-\arcsin\frac{2\Omega}{\mu\cos\alpha} 
\label{equilibria}
\end{equation}

corresponding to stable node $O=(\eta_O;0)$ and saddle $S=(\eta_S; 0)$ of the in-manifold scalar Eqn.~\ref{Sync2}. 
The states are born in a saddle-node bifurcation at the bifurcation curve

\begin{equation}
\mu_{sync}=\frac{2\Omega}{\cos\alpha},
\label{bif_curve}
\end{equation}

and they exist for all $\mu>\mu_{sync}$, where node $O$ transforms into a stable focus soon after the bifurcation 
at $\mu_{foc}=\frac{2\Omega}{\cos\alpha}\sqrt{1+\epsilon^{4}/64}$. 

The synchronous state we are looking for is given by the stable equilibrium
$O$. The in-manifold stability, however, does not guarantee its transverse
stability. Our numerics indicate, however, that the equilibrium $O$ is also stable
in the whole $N$-dimensional space of the original model Eqn.~\ref{sync1}.

We conclude that the synchronous state of the circle model (Eqn.~\ref{sync1}) is born in a
saddle-node bifurcation as the coupling strength $\mu$ exceeds the bifurcation
value $\mu_{sync}$ and it preserves the stability with further increase of
$\mu$. In the state, generators and consumers create two phase-synchronized
groups with the phase shift between them equal $\eta_O$. At the saddle-node
bifurcation curve $\mu=\mu_{sync}$, the phase shift is $\eta_O=\pi/2$ and it
decreases to zero as $\mu\rightarrow\infty$. Frequency $\Omega_O$ of the
synchronous state $O$ is obtained from Eqn.~\ref{sync1} as:

\begin{equation}
\Omega_O=-\frac{\mu\sin\alpha}{2\epsilon}(1-\sqrt{1-(\frac{2\Omega}{\mu\cos\alpha})^{2}})=-\frac{\mu(1-\cos\eta_O)}{2\epsilon}\sin\alpha.
\end{equation}

It follows therefore that synchronous frequency $\Omega_O$ is zero in the
lossless case $\alpha=0$, negative for $\alpha>0$ and positive for $\alpha<0$.

\subsection*{Solitary states in mean-field coupled circle model}

Solitary states can be derived analytically in the case of mean-field
approximation of the circle model, i.e. when $R=N/2$ in Eqn.~\ref{sync1}.
For convenience, we write the mean-field model in the standard normalized form

\begin{align}
\ddot \phi_i+ \epsilon \dot \phi_i &= (-1)^{i+1}\Omega +\frac{\mu}{N}\sum\limits_{j=1}^{N} \sin(\phi_j - \phi_i - \alpha)\;
\label{mean-field}.
\end{align}

Here parameter $\alpha$ can be positive or negative. Due to the assigned
symmetry, these two cases are reducing explicitly to each other by simple
transformation $\alpha \rightarrow -\alpha, ~ \Omega \rightarrow -\Omega,
~\phi_i \rightarrow -\phi_i$, at which the generators and consumers exchange
their network disposition. Therefore, the system dynamics appears to be
equivalent for $\alpha<0$ and $\alpha>0$ and if so, only one of the cases
should be derived. The symmetry property is also valid for the general, non mean-field 
form of the circle power grid model (Eqn.~\ref{sync1}), which is clearly
demonstrated by our simulations. However, it is not the case for non-symmetric
generator/consumer disposition as it is normally the case in real power grid
networks, e.g. in the Scandinavian power grid.

The synchronous state $O$ of the mean field-model Eqn.~\ref{mean-field} is obtained by
the same procedure as in the previous section applied to the two-group
representation Eqn.~\ref{two-group} including generators $\theta_i$ and consumers
$\psi_i$:

\begin{align}
\ddot \theta_i+ \epsilon \dot \theta_i &= +\Omega +\frac{\mu}{N}\sum\limits_{j=1}^{N/2}(\sin(\theta_j - \theta_i - \alpha)+\sin(\psi_j - \theta_i - \alpha)),\nonumber\\
\ddot \psi_i+ \epsilon \dot \psi_i &= -\Omega +\frac{\mu}{N}\sum\limits_{j=1}^{N/2} (\sin(\psi_j - \psi_i - \alpha)+\sin(\theta_j - \psi_i - \alpha)).\nonumber\\
\end{align}

So, it has the form 

\begin{equation}
O: ~\theta_1=\dots=\theta _{N/2}\equiv\theta, ~~\psi_1=\dots=\psi _{N/2}\equiv\psi
\end{equation}

with a constant phase shift $\eta_O=\theta-\psi$ between generators and consumers, where $\eta_O=\arcsin\frac{2\Omega}{\mu\cos\alpha}$; exists and is stable for all $\mu$ above the bifurcation curve as in \ref{bif_curve}. 

Beyond the the existence of synchronous state $O$ which is phase and frequency
synchronized, numerous solitary states of different configurations are
typically arising in the mean-field model (Eqn.~\ref{mean-field}). They
consist normally of several phase clusters, each rotating with its own
frequency, manifesting in such a way a phenomenon of {\it frequency
clustering}. This peculiar phenomenon differs from usual phase clustering
well-known in (pure) phase oscillatory networks, cr. standard Kuramoto model.
De facto, due to the presence of inertia, the system dynamics becomes much
more involved. The main peculiarity is an enormous multistability, i.e., when
the number of co-existing stable solitary and other states can grow
exponentially with the system size $N$, moreover, in an essential domain of
the system parameters. If so, {\it spatial chaos}
\cite{elphick1987nature,NIZHNIK2002,Omelchenko2011} becomes an inherent
characteristic of the model. We are coming apparently to a new reality, still
not comprehended, in the field of coupled oscillators with inertia.

The simplest but not trivial solitary-type behaviour is supplied by a
1-solitary state, where just one generator or consumer- let for definiteness
it be a generator $\theta_1$ - splits off from the others, which are frequency
synchronized consisting of two phase clusters: one of remaining generators and
the other of all consumers, i.e.,

\begin{equation}
\Phi_{1-sol}: ~\theta_1\equiv\gamma; ~\theta_2=\dots=\theta _{N/2}\equiv\theta, ~\psi_1=\dots=\psi _{N/2}\equiv\psi.
\label{1-sol}
\end{equation}

\begin{figure}
\includegraphics[scale=0.7]{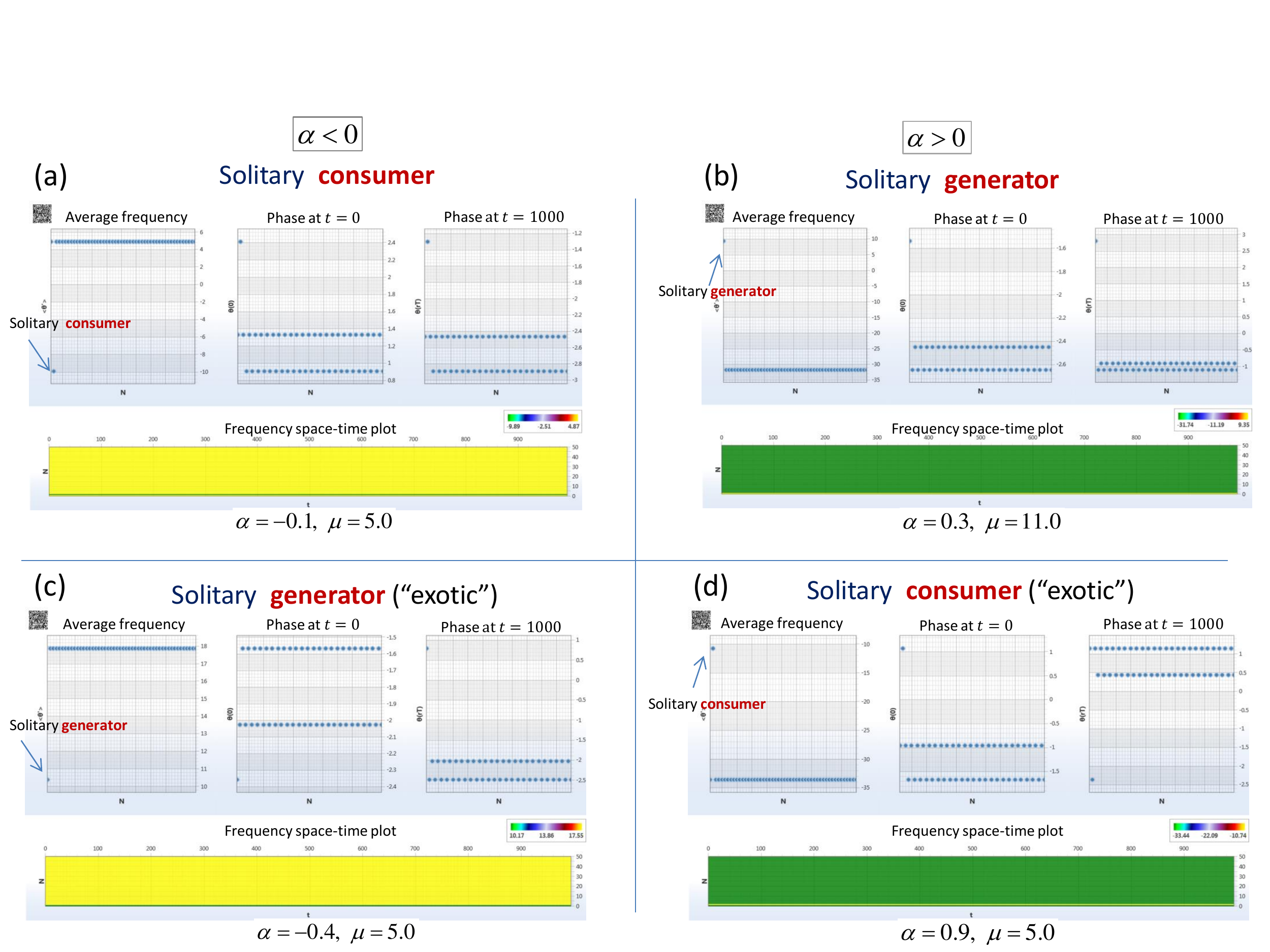}
\caption{(colour online). 1-solitary states in mean-field coupled model
(Eqn.~\ref {mean-field}). Mean frequency, and two phase snapshots are shown
above, and frequency space-time plot is below. All simulation starting from
random initial conditions, the behaviour is shown after transient $T=10^4$
time units. Parameters as in the main text.}
\label{fig1S}
\end{figure}

Examples of four different 1-solitary states for the mean-field model
(Eqn.~\ref {mean-field}) are illustrated in Figure~\ref{fig1S}, for solitary
generator (b,c) and consumer (a,d). Due to the model symmetry, the case with
solitary consumers can be reduced to the generator one by simple
parameter/variable change $\alpha \rightarrow -\alpha, ~ \Omega \rightarrow
-\Omega, ~\phi_i \rightarrow -\phi_i$. We skip it in the section and
concentrate further on the states with a solitary generator as in
Fig.~\ref{fig1S}.

As it is illustrated in Fig.~\ref{fig1S}, there are two types of
1-solitary states in which a generator is detached. Actually, what is happened
in both cases, the generator continues to rotate close to its eigenfrequency
$\Omega=10$ in the situation when all others units synchronize with smaller
(b) or larger (c) mean frequency. The first situation (b) is maybe considered
a "normal", as the desirable synchronous frequency should naturally be less
then those of the generators, and it equals zero if $\alpha=0$ and deviates to
negative values as $\alpha>0$. Another 1-solitary state shown in (c) obeys a
counter-intuitive property: the mean frequency of the synchronous cluster appears
to be larger then 10, we call this state an ''exotic'', and it exist for
negative $\alpha$ only.

\begin{figure}
\includegraphics[scale=0.7]{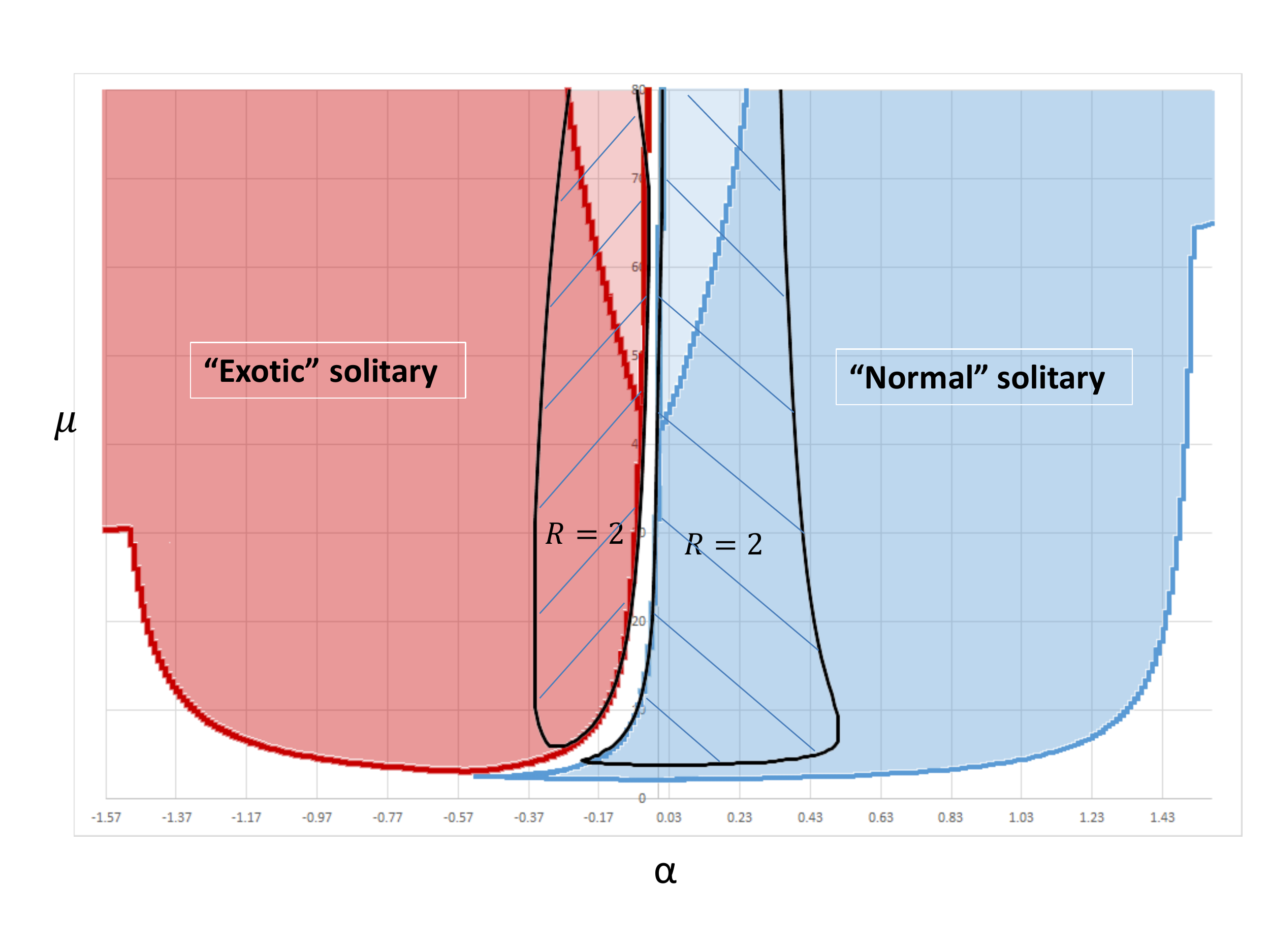}
\caption{(colour online). Parameter regions for both, ``normal'' and ``exotic''
1-solitary states in the mean-field coupled model Eqn.~\ref{mean-field} (coloured)
and the respective locally coupled model with coupling radius $R=2$ (dashed).
Parameters as in the main text.}
\label{fig2S}
\end{figure}

The parameter region for both, ''normal'' and ''exotic'' 1-solitary states (with a
generator detached) are shown in Fig.~\ref{fig2S}. Analogous regions for the
case with a consumer detached (see examples in Fig.~\ref{fig1S} (a,d)) are
symmetrically disposed with respect to the $\alpha=0$ axis.

\subsection*{Reduction to Solitary Manifold} 

In the new variables $(\gamma, \theta, \psi)$ for 1-solitary state (Eqn.~\ref{1-sol})
the mean-field model (Eqn.~\ref{mean-field}) consists of only three equations:

\begin{align}
\ddot \gamma+ \epsilon \dot \gamma &=+\Omega +\frac{\mu}{N}((\frac{N}{2}-1)\sin(\theta-\gamma- \alpha)+\frac{N}{2}\sin(\psi - \gamma - \alpha)+\sin(-\alpha)),\nonumber\\
\ddot \theta+ \epsilon \dot \theta &= +\Omega +\frac{\mu}{N}(\frac{N}{2}\sin(\psi - \theta-\alpha)+\sin(\gamma - \theta - \alpha)+(\frac{N}{2}-1)\sin(-\alpha)),\nonumber\\
\ddot \psi_i+ \epsilon \dot \psi_i &= -\Omega +\frac{\mu}{N}((\frac{N}{2}-1)\sin(\theta-\psi-\alpha)+\sin(\gamma-\psi-\alpha)+\frac{N}{2}\sin(-\alpha)).\nonumber\\
\label{1-sol_system-6}
\end{align}

This is a reduced model of three coupled virtual ''pendula'' governing the
dynamics in the respective 6-Dim solitary manifold. The pendula are linear but the
coupling terms, however, are strongly non-linear (contain sinusoidal
functions). This results in complex dynamical behaviour characterized by
multiple solitary states and chaos. As coupling terms depend only on the phase
differences, the system dimension can be further reduced by introducing phase
difference variables $\nu=\gamma-\psi, ~~ \eta=\theta-\psi$. In the new
variables, only two equations are left:

\begin{align}
\ddot \nu+ \epsilon \dot \nu &=2\Omega +\frac{\mu}{2}((1-\frac{2}{N})\sin\alpha-\sin(\nu+\alpha)+\frac{2}{N}\sin(\nu-\alpha)+(1-\frac{2}{N})(\sin(\eta-\nu-\alpha)-\sin(\eta-\alpha))),\nonumber\\
\ddot {\eta}+ \epsilon \dot {\eta} &=2\Omega+\frac{\mu}{N}\sin\alpha-\frac{\mu}{2}(\sin(\eta+\alpha)+(1-\frac{2}{N})\sin(\eta-\alpha))+\frac{\mu}{N}(\sin(\nu-\eta-\alpha)-\sin(\nu-\alpha)).\nonumber\\
\label{1-sol_system-4}
\end{align}

This is a system of only two coupled ''pendula'', maintaining on a 4-Dim
invariant manifold of the whole $N$-dimensional phase space. It is complicated
for analytic study, as each pendulum as well as all coupling terms are non-linear. 
However, the model is low-dimensional so it can be easily derived in a
standard way with use of numerical simulations to identify 1-solitary states
for the $N$-dimensional mean-field model (Eqn.~\ref{mean-field}), up to the
thermodynamic limit $N\rightarrow\infty$. Note, that all in-manifold regimes
which develop within the framework of the reduced system
(Eqn.~\ref{1-sol_system-4}) should be tested on the transverse stability with
respect of out-of-manifold perturbations, to guarantee the existence in the
whole $N$-dimensional model.

This approach is also applicable to $k$-solitary states, with any $k=2,3..$, of the form
\begin{equation}
\Phi_{k-sol}: ~\theta_1=\dots=\theta_k\equiv\gamma; ~\theta_{k+1}=\dots=\theta _{N/2}\equiv\theta, ~\psi_1=\dots=\psi _{N/2}\equiv\psi.
\label{k-sol}
\end{equation}

Reduced low-dimensional systems for the $k$-solitary states can be obtained in
a similar way as in the $k=1$ case above. They are slightly different with the
system parameters however, their dimensions are the same. Note finally, that the
behaviour of solitary states becomes even more puzzled if the synchronized
elements, variables $\theta_i$ and $\psi_i$ in (Eqn.~\ref{k-sol}), split off by
more then two clusters. Above reduced procedure still works, however, the
dimension of the ''solitary'' manifold becomes larger then $4$.  Nevertheless,
the study in this case is yet much simpler then the original $N$-dimensional
model.

\subsection*{Homoclinic Bifurcation of Solitary Oscillations}

As can be seen for instance in Fig.~\ref{fig4}, solitary oscillations cease to exist beyond 
a critical value for the losses $\alpha$ (at fixed coupling strength). 
Our numerical studies give a strong indication that the underlying mechanism is a homoclinic 
bifurcation of the solitary limit cycle towards synchronisation.
Furthermore, this is also suggested by our analogy with the infinite-bus-model.
The numerical procedure is as follows:
The period of the solitary oscillation is estimated as the inverse time average of 
the phase velocity $\omega_s$ at the solitary node with index $s$:

\begin{equation}
  \tau = \frac{2 \pi}{\overline{\omega_s}} = 2\pi T \left( \int\limits_0^T~dt\; \omega_s (t)\right)^{-1}
\end{equation}
  
for some large enough $T$ such that we can be sure to measure several periods.
Varying the phase lag $\alpha$ as the control parameter,  we used the period
$\tau$ estimated at one parameter value to  set $T=10\tau$ for the next
(updated) parameter value. The numerical integration is performed using
SciPy's \verb+CVode+ solver with step size $10^{-3}$.

In case of a homoclinic bifurcation, it is expected that the period $\tau$
scales logarithmically with the parameter in proximity to the bifurcation
point $\alpha_c$, i.e. $\tau \simeq \log \vert \alpha_c - \alpha\vert $ 
(see for instance \cite{Strogatz2018}).

\begin{figure}[!ht]]
   \begin{subfigure}[c]{.4\columnwidth}
   \centering
   \caption{}
    \includegraphics[width=\textwidth]{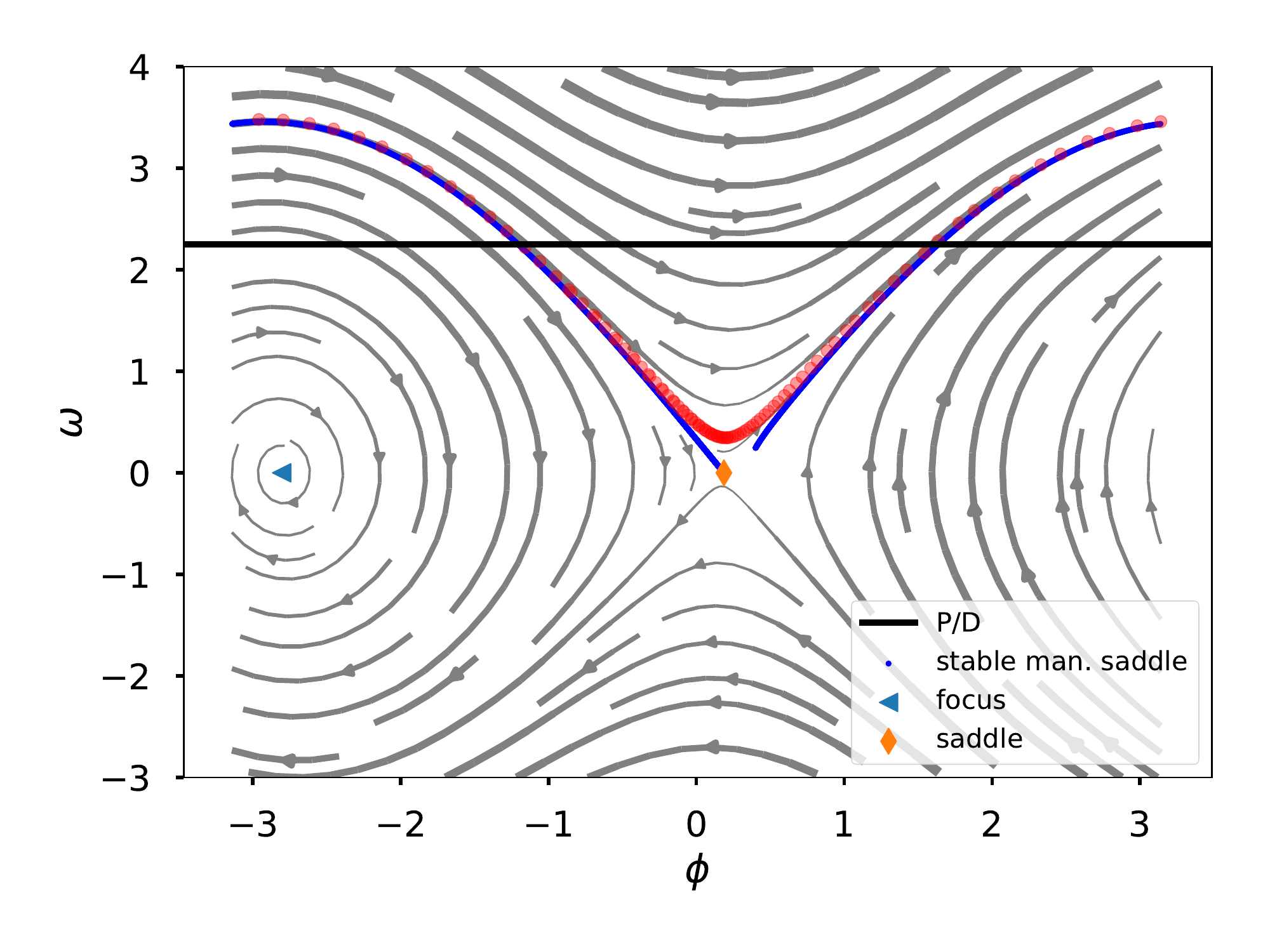}
    \label{fig:bifurcation-a}
  \end{subfigure}
  \begin{subfigure}[c]{.28\columnwidth}
  \centering
  \caption{}
    \includegraphics[width=\textwidth]{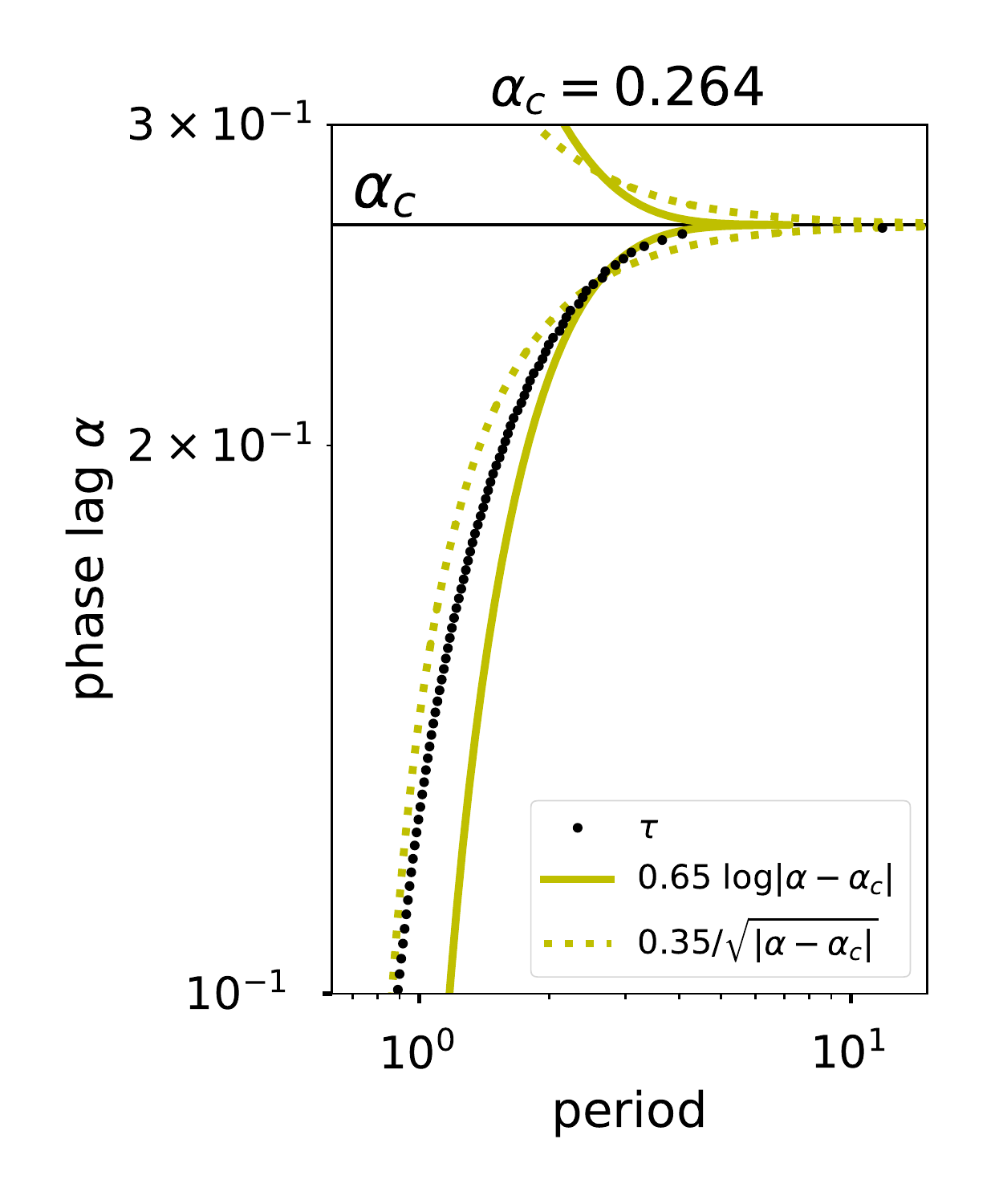}
    \label{fig:bifurcation-b}
  \end{subfigure}
  \begin{subfigure}[c]{.28\columnwidth}
  \centering
  \caption{}
    \includegraphics[width=\textwidth]{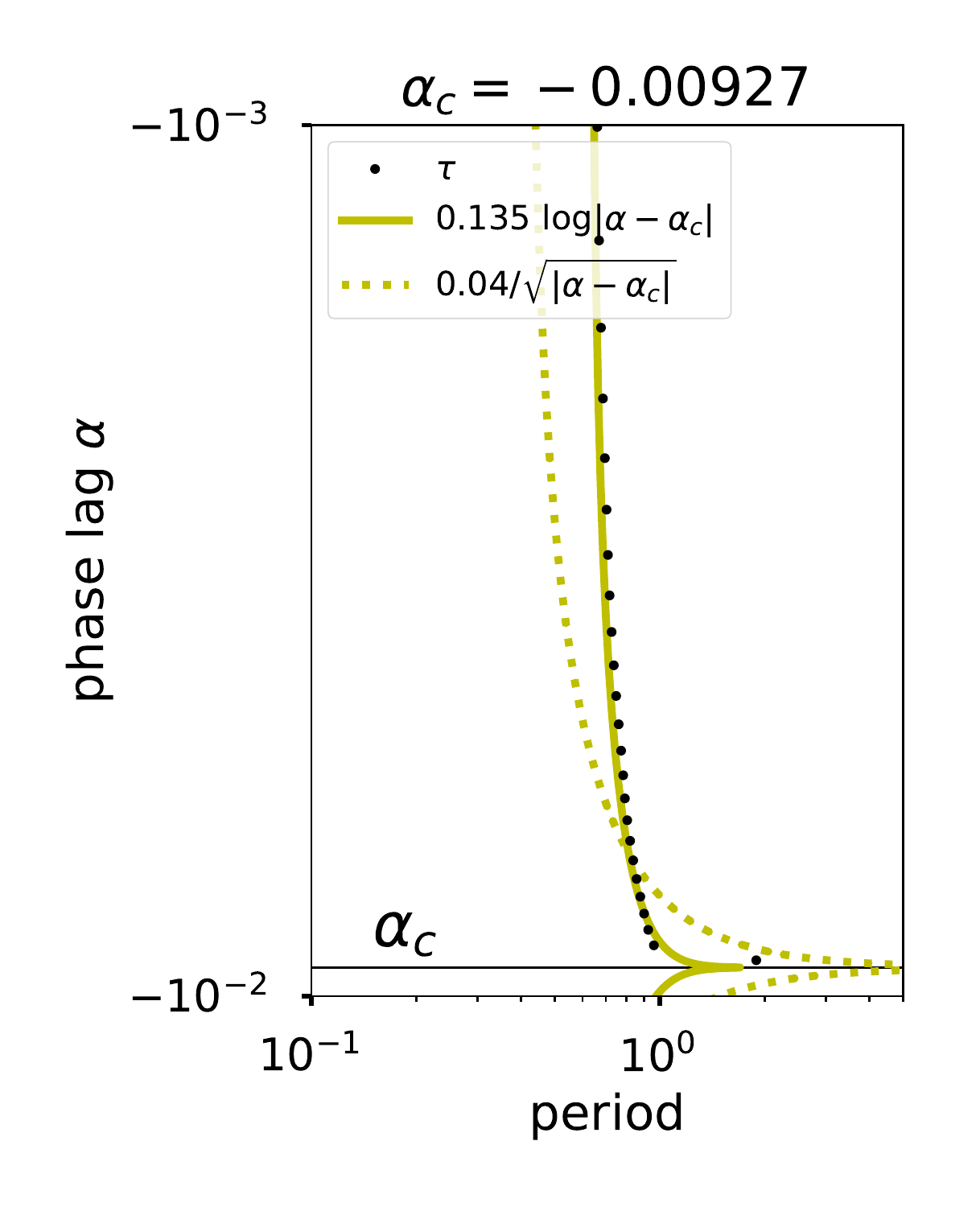}
    \label{fig:bifurcation-c}
  \end{subfigure}
  \caption{\textbf{Bifurcation analysis.}
  \textbf{a} Phase space of the infinite bus model $P=1$, $D=0.1$, $K=2$,
$H=1$. The positions of the stable focus,   the saddle and its stable manifold
are indicated. The horizontal black line indicates the value of
$\omega_{lc}^\prime$.   See the text for further discussion.
  \textbf{b} Bifurcation analysis of the infinite bus model with above
  parameters.
  \textbf{c} Bifurcation analysis of the circle model with standard parameters
  and control.
  \textbf{b},\textbf{c} The yellow lines depict logarithmic and square root
scaling laws with suitable parameters as visual guidance.
  }
 \label{fig:bifurcation}
\end{figure}
 
Consider the phase portrait of the infinite bus model ($P=1$, $D=0.1$, $K=2$,
$H=1$)  close to the bifurcation point at $\alpha_c\approx 0.263$ pictured in
Fig.~\ref{fig:bifurcation-a}. The red circles correspond to 100 initial
conditions equally spaced on the limit cycle. Firstly, the locus of the limit
cycle is close to the stable manifold of the saddle point (determined by
backward integration) before they merge into a homoclinic orbit at $\alpha_c$.
Secondly, the distribution of points clusters near the saddle, indicating an
increase  of $\tau$ close to $\alpha_c$ due to a slowing down of trajectories.
The numerically determined values of $\tau$ are given in
Fig.~\ref{fig:bifurcation-b}. As a guidance to the eye, the solid yellow line
depicts a logarithmic scaling as expected for the homoclinic bifurcation in
the infinite bus model.  For comparison, the dashed yellow line gives a square
root scaling in the  case of an infinite-period bifurcation. It appears that
close to $\alpha_c$ the logarithmic function is a better candidate for the
scaling $\tau (\alpha)$, the onset of the divergence is much steeper than
expected for the square root case. Analogously, $\tau (\alpha)$ is estimated
for the circle model, here for the standard  parametrization and control. From
Fig.~\ref{fig4a} at $K=6$, one expects a bifurcation  at slightly negative
$\alpha_c\approx -0.009$ from normal solitary oscillations  to
synchronisation. Repeating our analysis from the infinite bus model for the
circle indeed strongly suggests a homoclinic bifurcation
(Fig.~\ref{fig:bifurcation-c}). In summary, our numerical investigations and
the analogy to the infinite bus approximation  support the hypothesis that the
global bifurcation curves in the parameter plane of the  circle model are
homoclinic.

\subsection*{Additional Parameter Plane Studies}

\begin{figure}[!ht]
\begin{subfigure}[b]{0.32\columnwidth}
    \caption{Numerical continuation}
    \includegraphics[width=\textwidth]{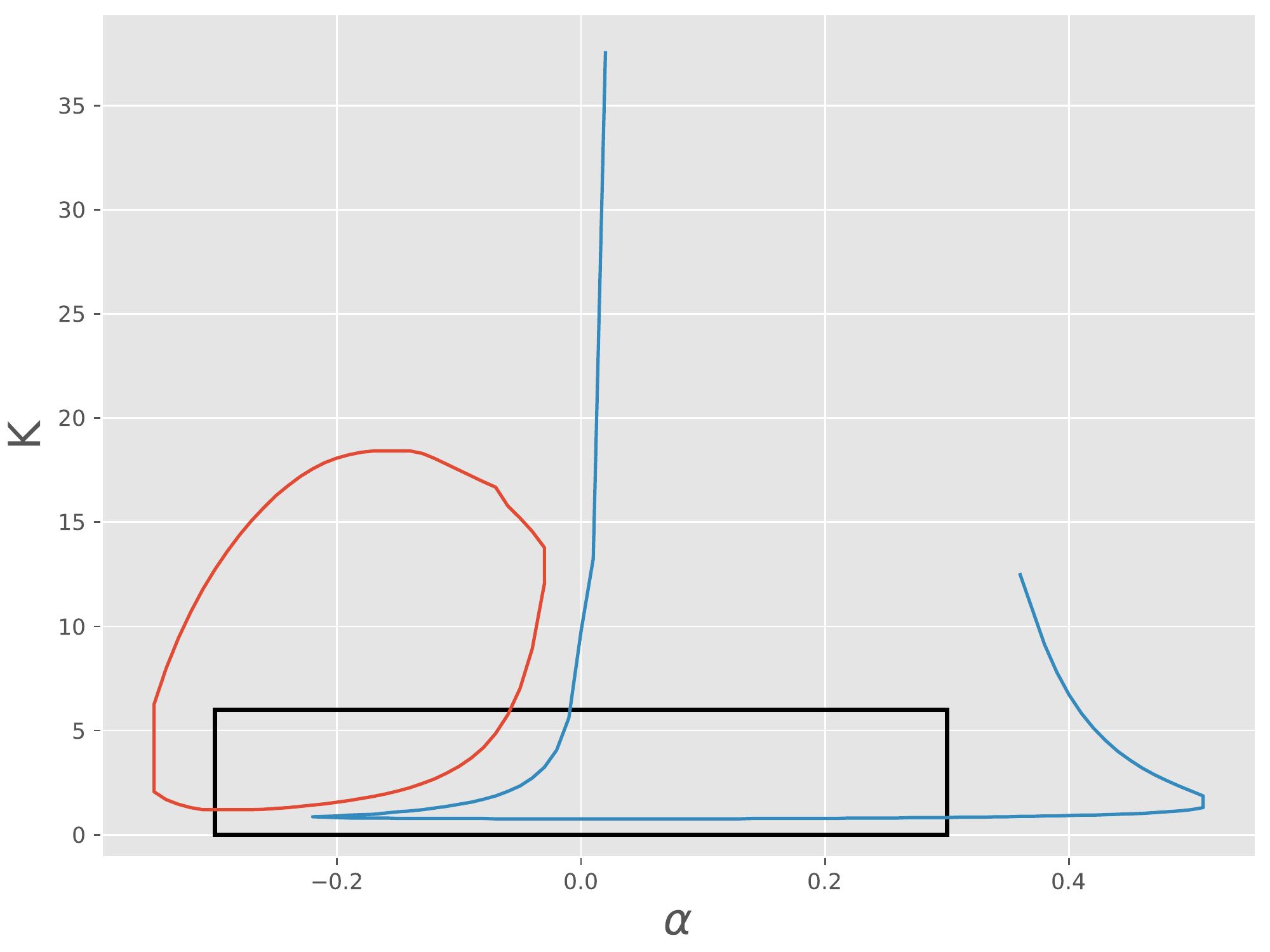}
  \end{subfigure}
   \begin{subfigure}[b]{0.32\columnwidth}
    \caption{global BS bimodal}
    \includegraphics[width=\textwidth]{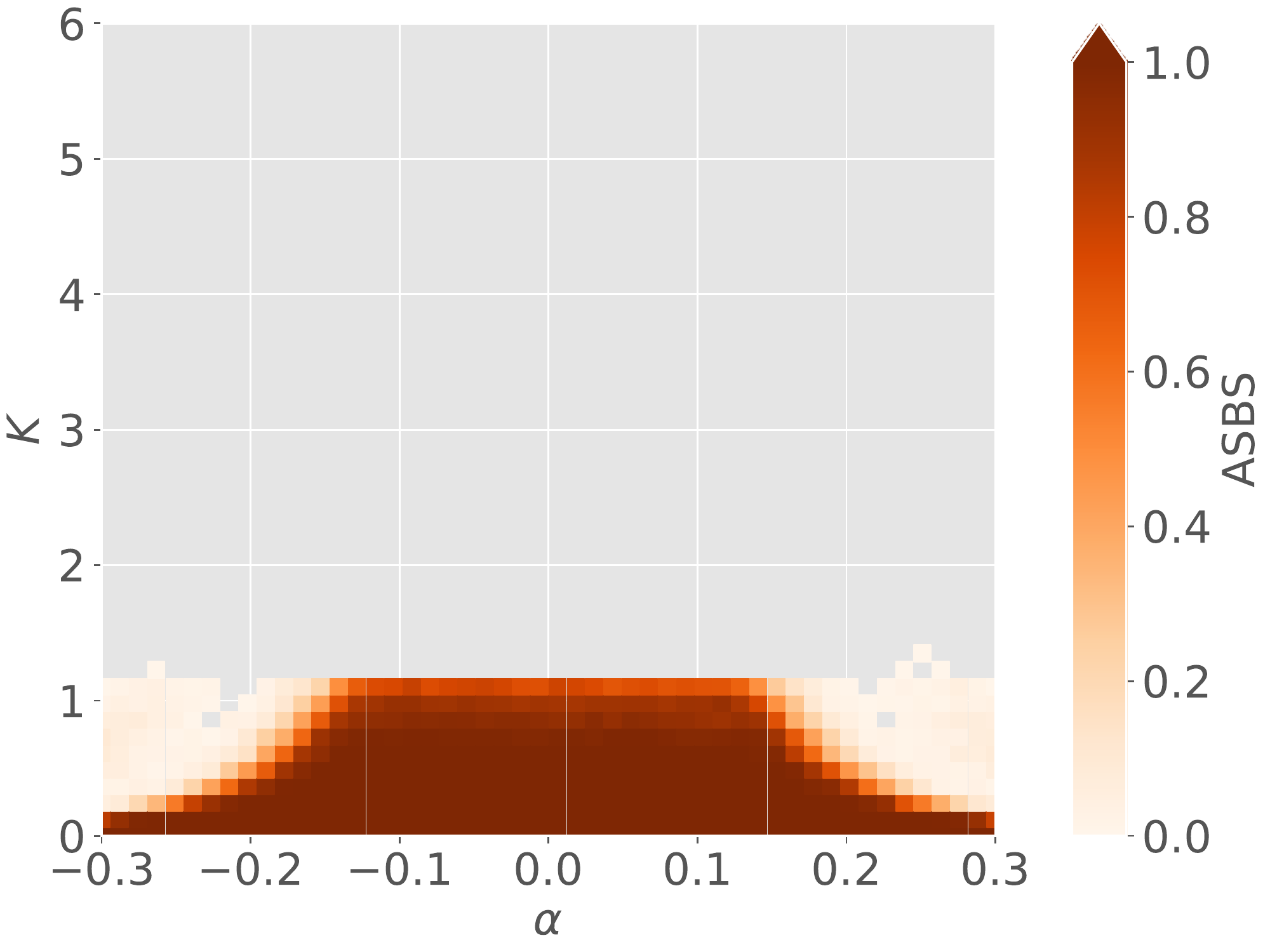}
  \end{subfigure}
   \begin{subfigure}[b]{0.32\columnwidth}
    \caption{ASBS synchronisation}
    \includegraphics[width=\textwidth]{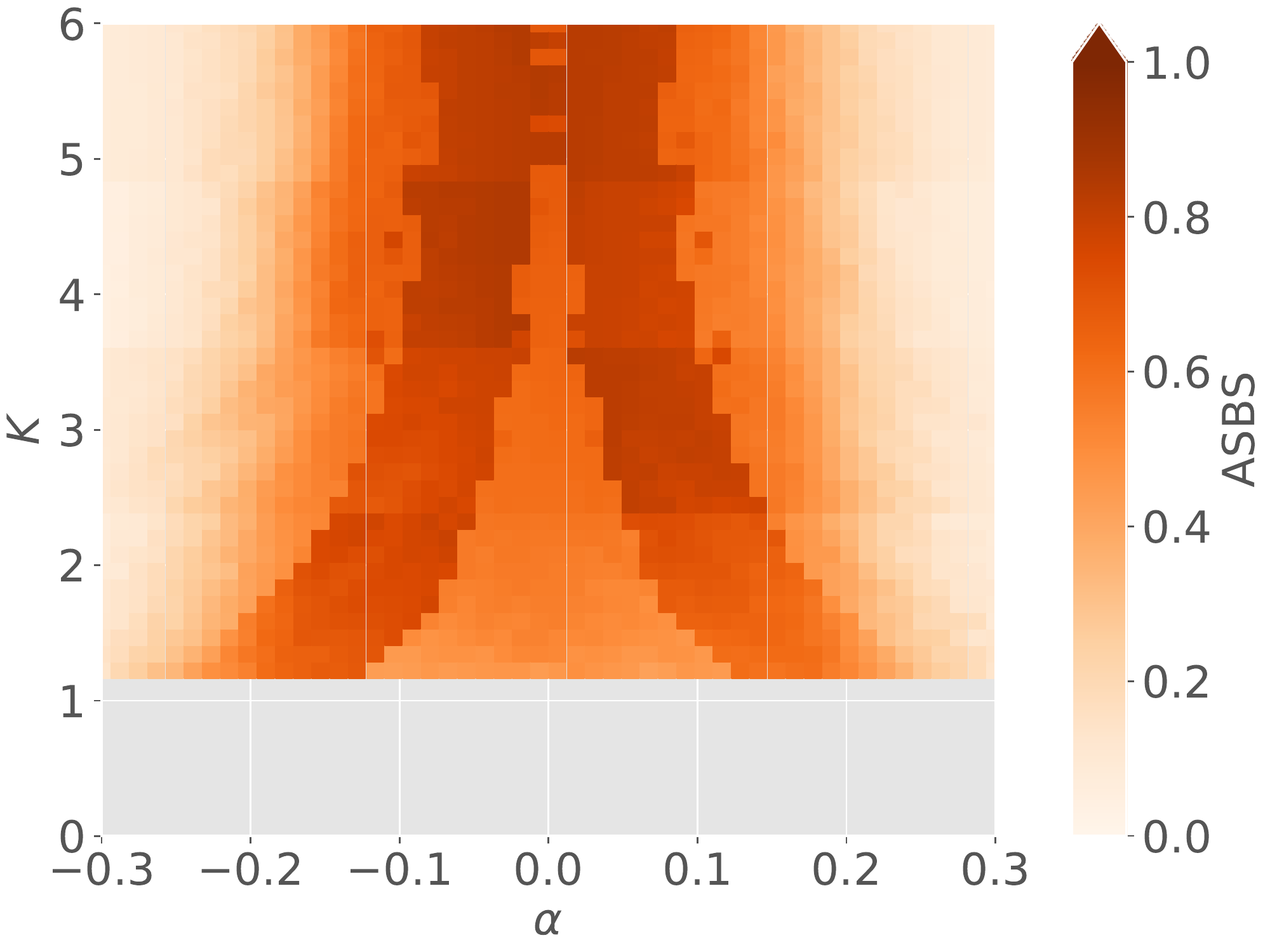}
  \end{subfigure}
  \caption{\textbf{Existence regions.} Simulation of the circle model $R=2$ with standard parametrization and control.
  \textbf{a} Boundary of the solitary existence region from Fig.~\ref{fig2S} with a box indicating the extent of our 
  numerical study in \textbf{b}/\textbf{c} as well as Figs.~\ref{fig4} and \ref{fig4_alt}.
  \textbf{b} Global basin stability estimate for the bimodal state.
  \textbf{c} Average single node basin stability estimate for the synchronous regime.}
 \label{fig:asbs-further}
\end{figure}

\begin{figure}[!ht]
   \begin{subfigure}[b]{0.48\columnwidth}
    \caption{Producer, standard parametrization}
    \includegraphics[width=\textwidth]{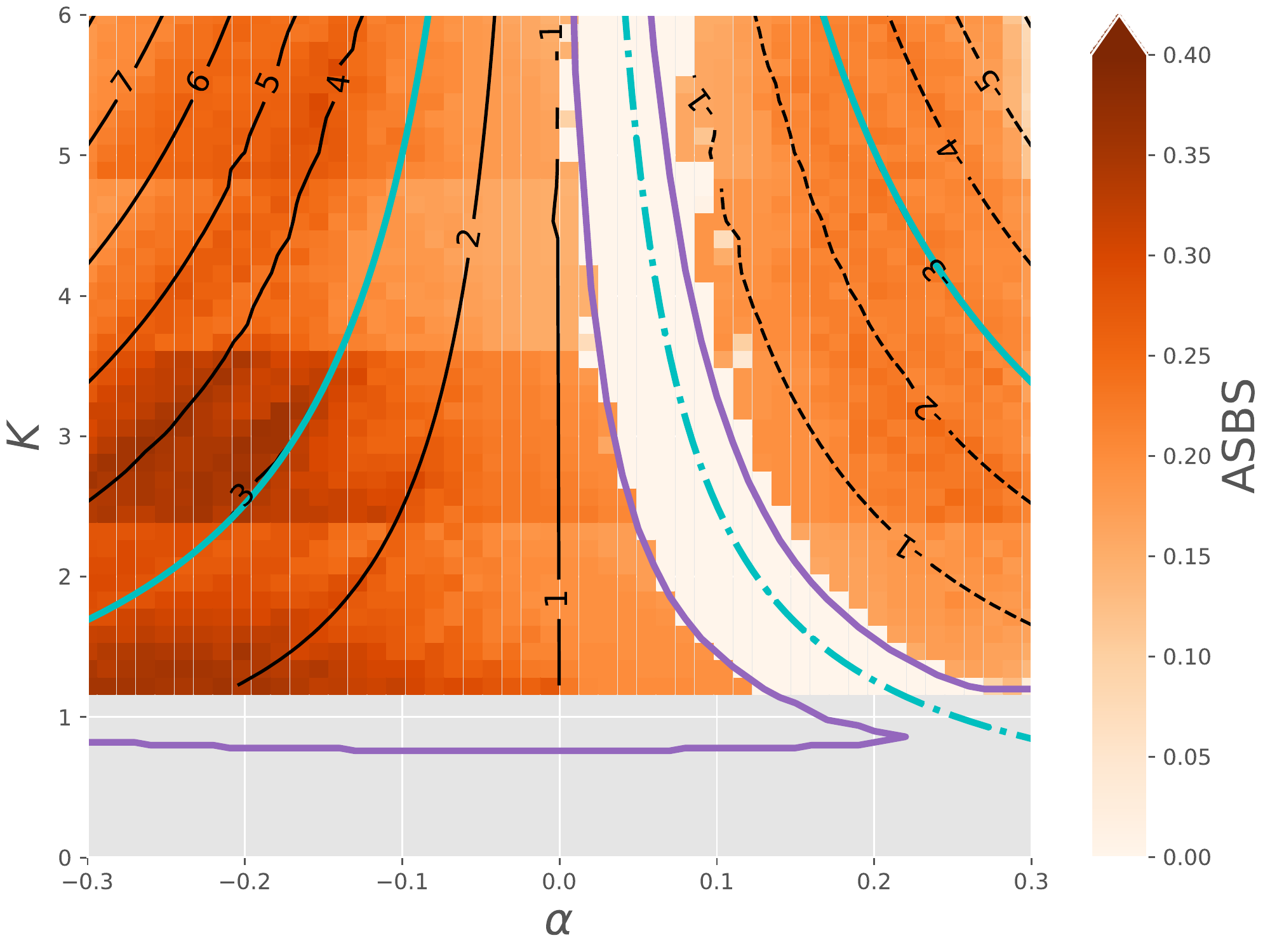}
  \end{subfigure}
   \begin{subfigure}[b]{0.48\columnwidth}
    \caption{Consumer, enhanced control} 
    \includegraphics[width=\textwidth]{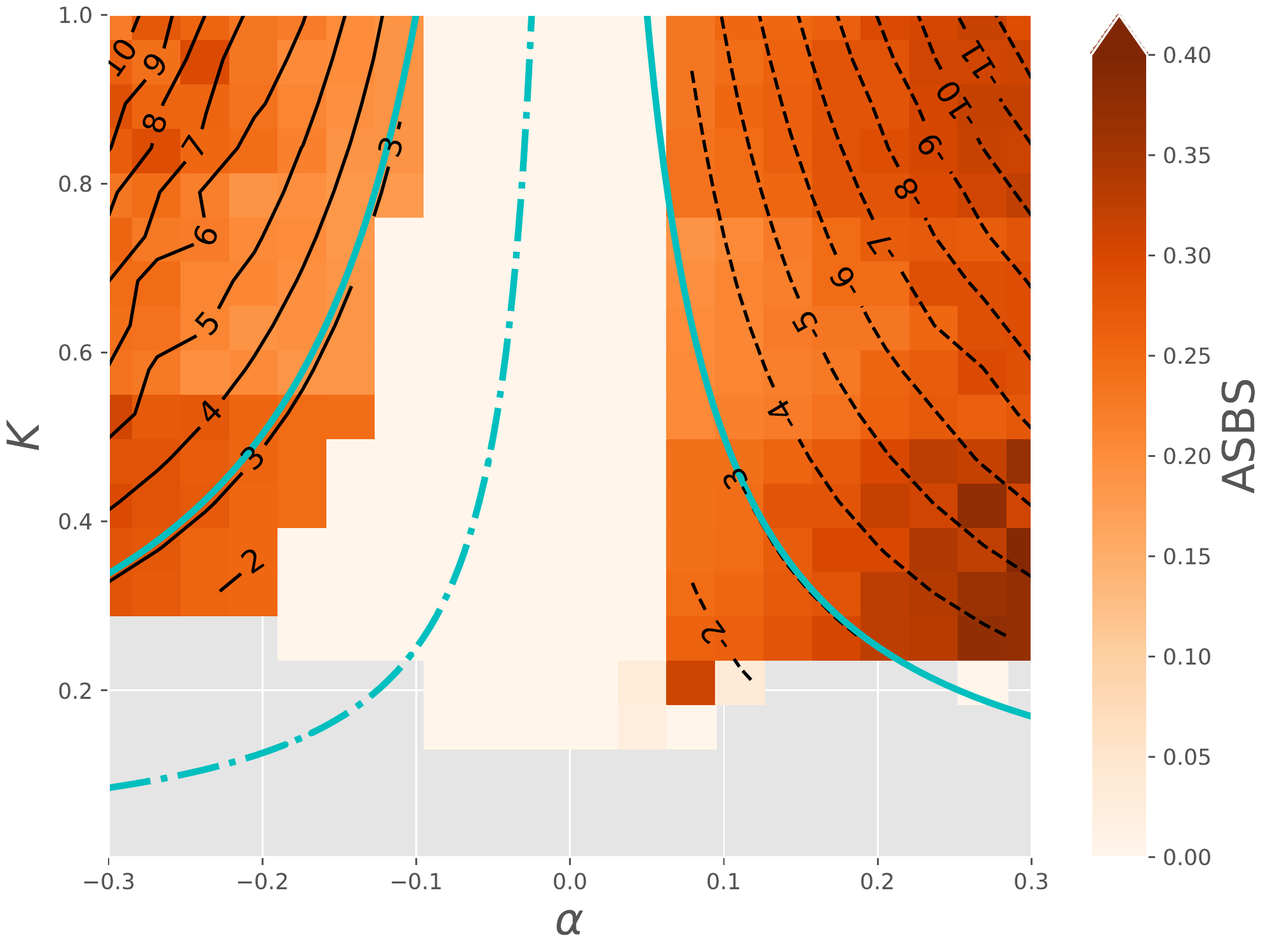}
  \end{subfigure}
  \caption{\textbf{Existence regions.} \textbf{a},\textbf{b} As Fig.~\ref{fig4}, with producer/conusmer exchanged.}
 \label{fig4_alt}
\end{figure}

\subsection*{Additional Basin Stability Studies}

The simulation results analysed with the sign opposed exotic solitaries marked
separately. For both, the average single node basin stability (ASBS) and the
global basin stability (global BS) the final state is clustered according to
frequency. The mean and maximum observed number of frequency clusters is shown
in the second column, the total number of oscillators outside the largest
cluster in the third column.

\begin{figure}[!ht]
  \caption{Scandinavian topology, strong control, standard coupling}
  \begin{subfigure}[b]{0.26\textwidth} 
    \caption{ASBS}
    \includegraphics[width=\textwidth]{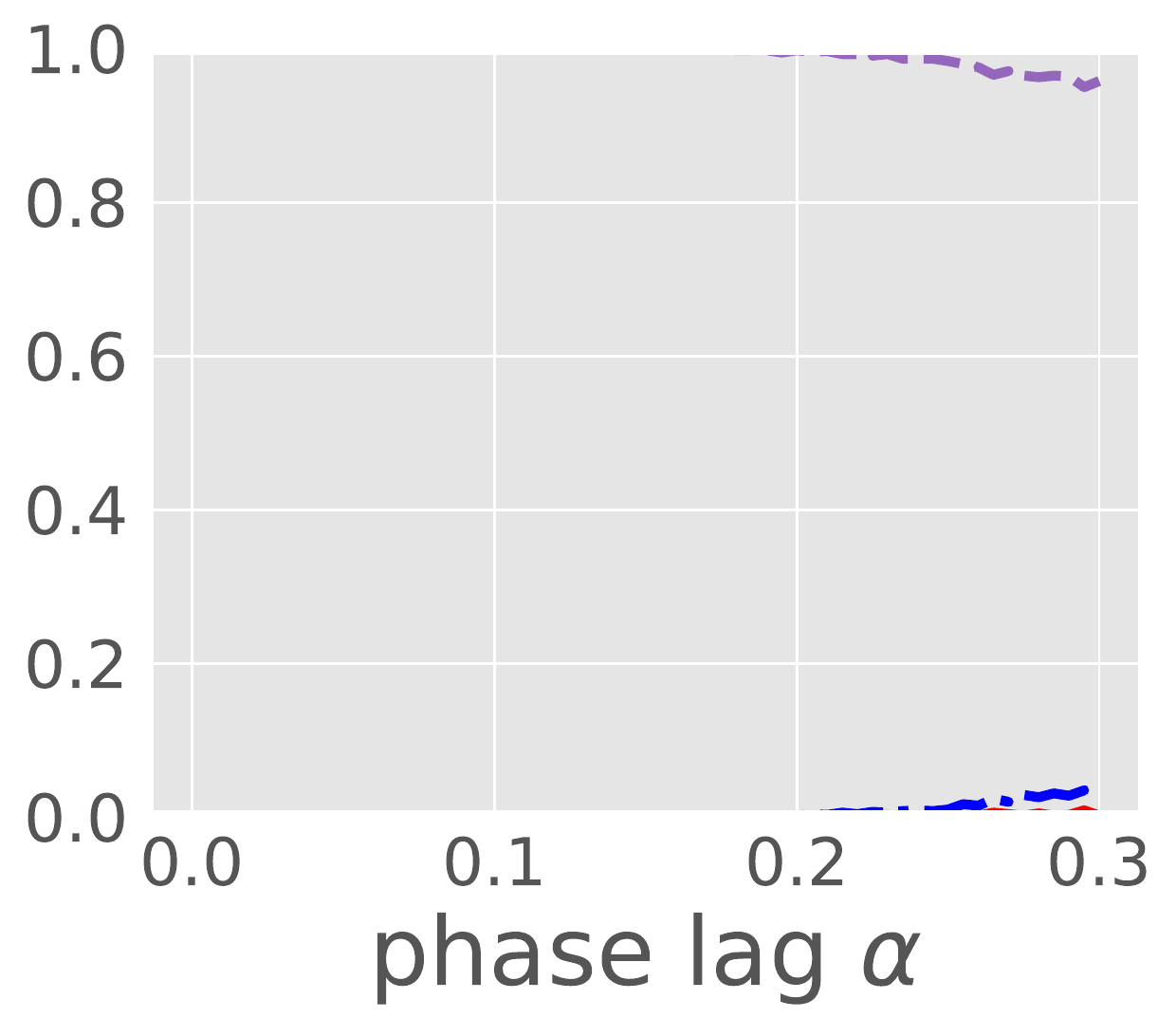}
  \end{subfigure}
  \begin{subfigure}[b]{0.26\textwidth}
    \caption{Expected no. of frequency clusters}
    \includegraphics[width=\textwidth]{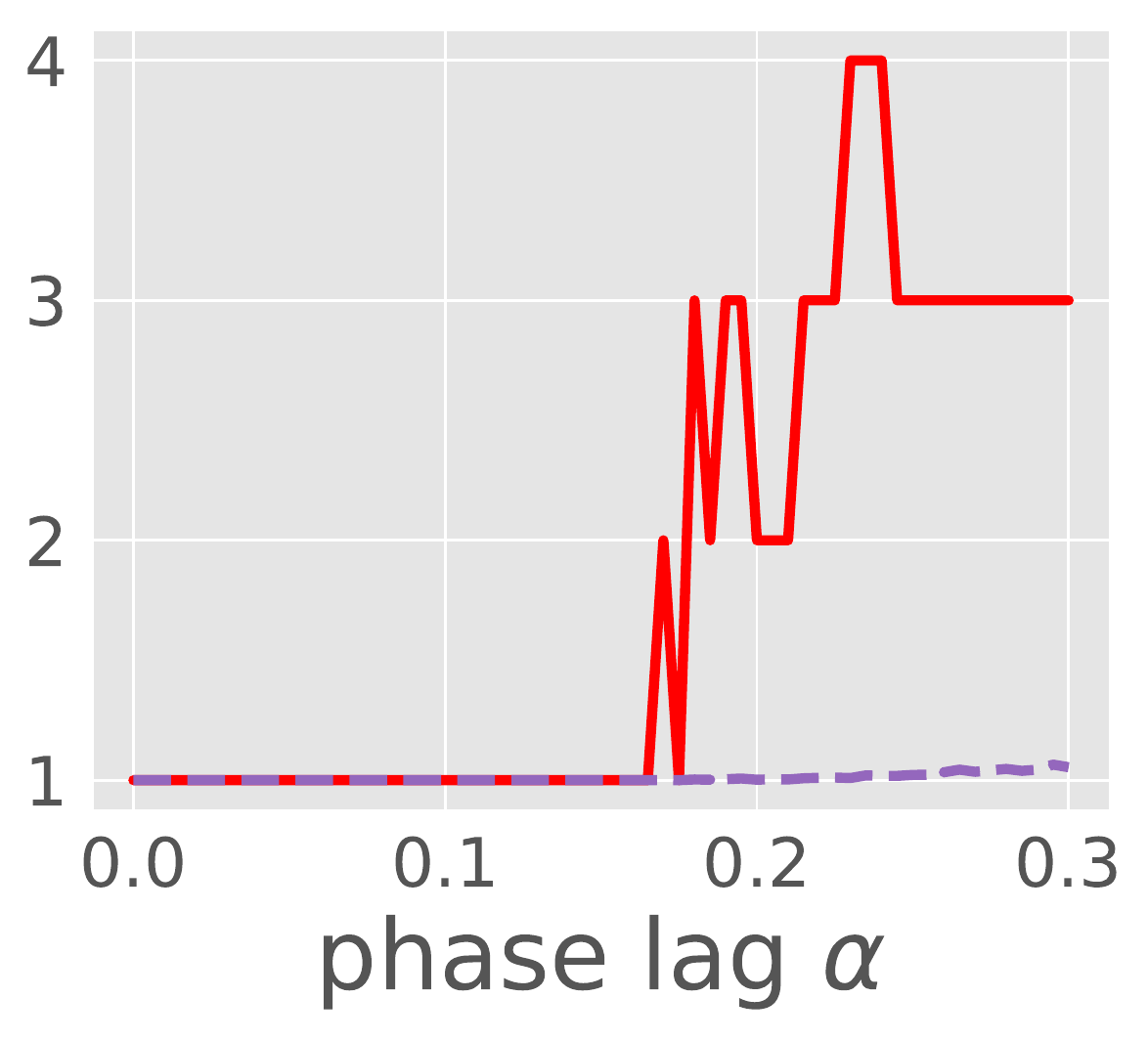}
  \end{subfigure}
  \begin{subfigure}[b]{0.26\textwidth}
    \caption{Expected no. of desync oscillators}
    \includegraphics[width=\textwidth]{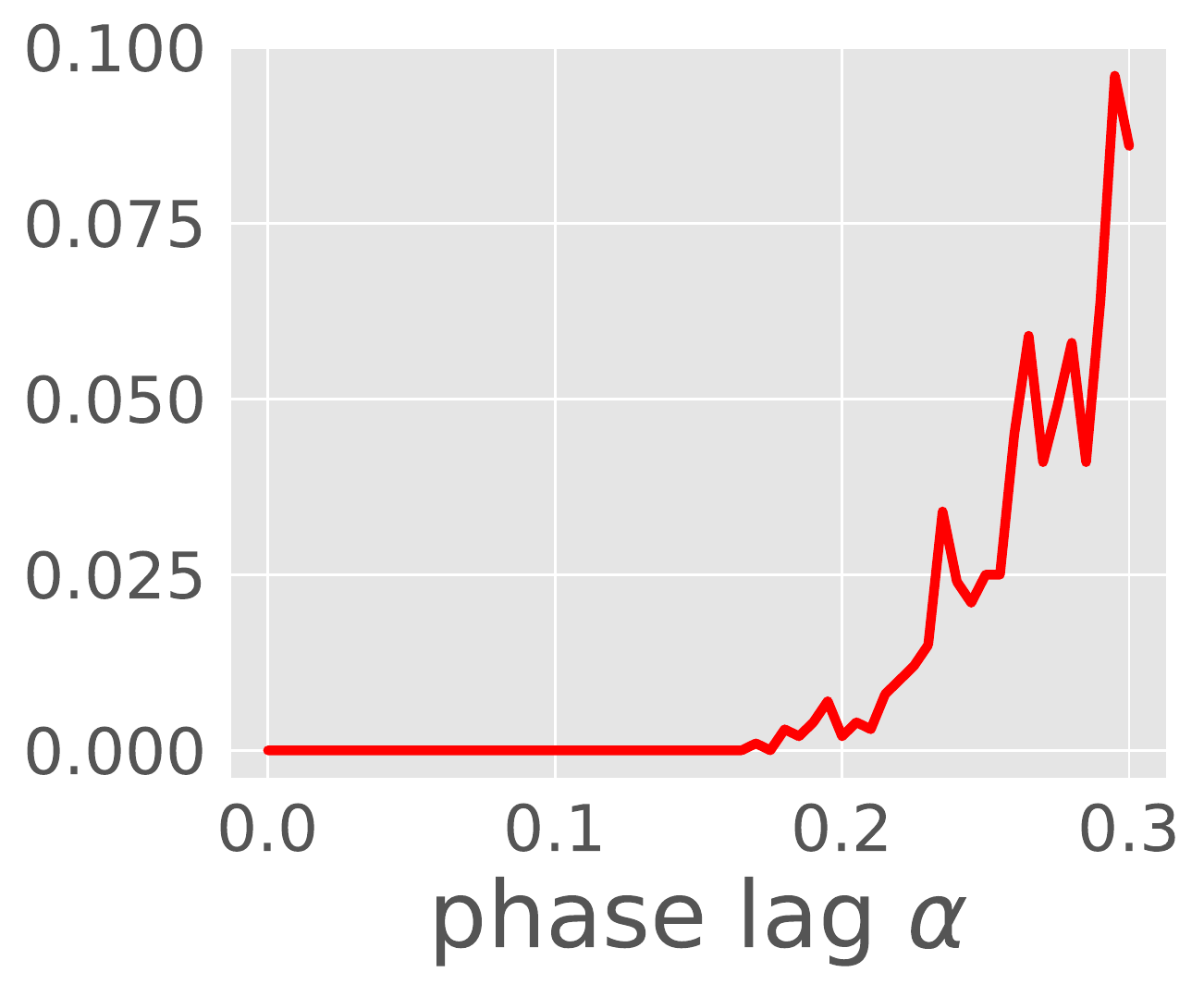}
  \end{subfigure}

  \begin{subfigure}[b]{0.26\textwidth}
    \caption{global BS}
    \includegraphics[width=\textwidth]{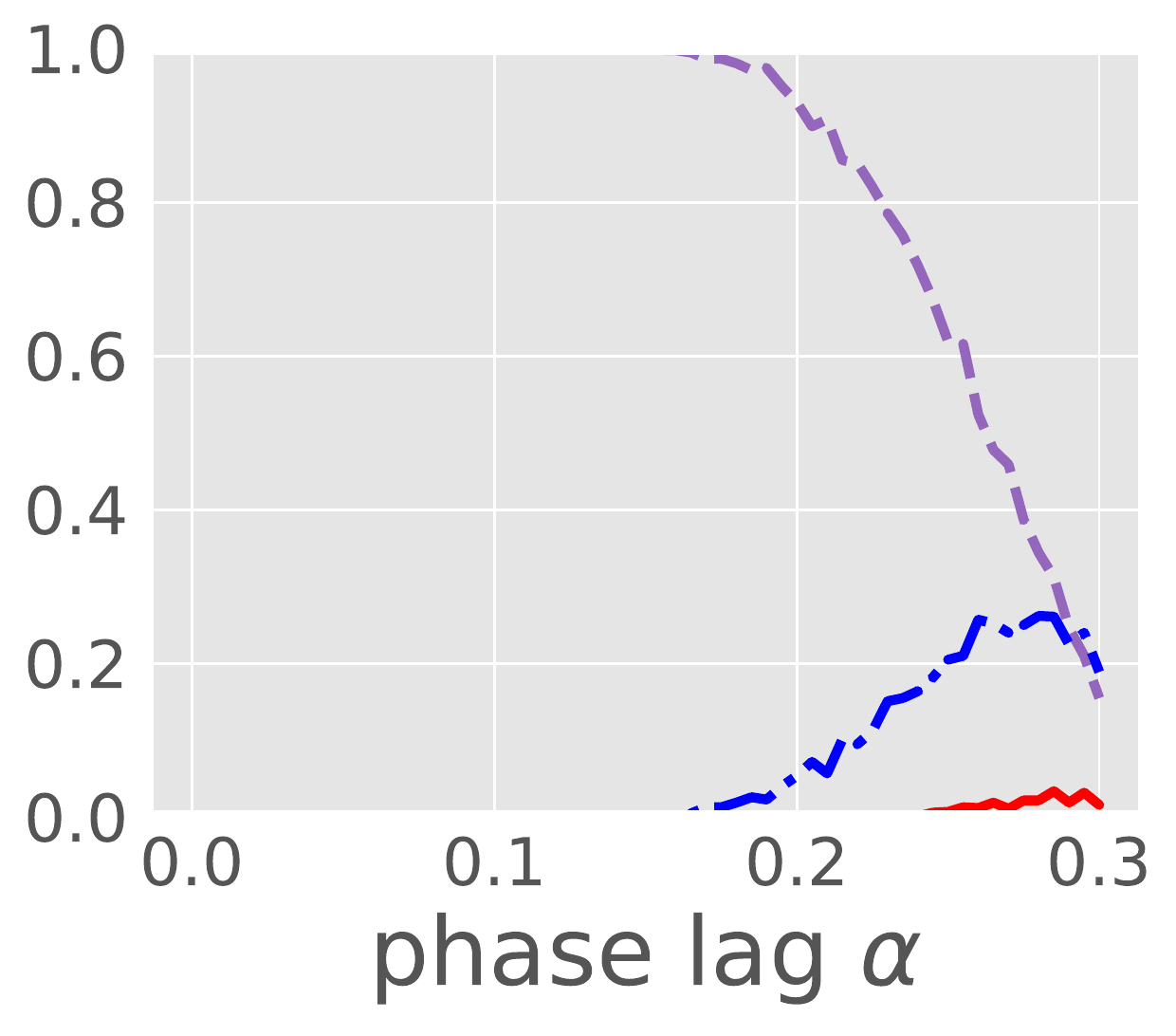}
  \end{subfigure}
  \begin{subfigure}[b]{0.26\textwidth}
    \caption{Expected no. of frequency clusters}
    \includegraphics[width=\textwidth]{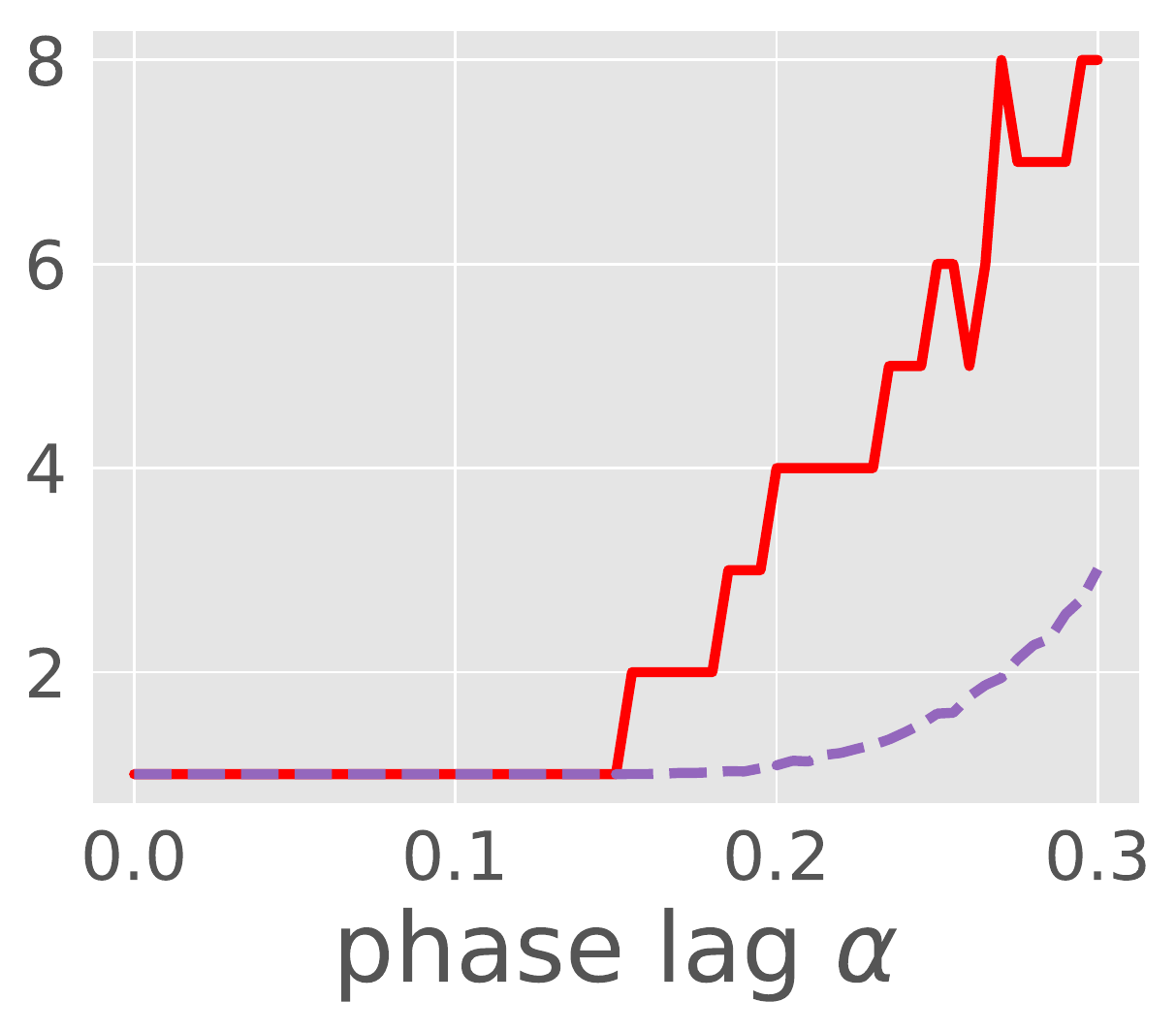}
  \end{subfigure}
  \begin{subfigure}[b]{0.26\textwidth}
    \caption{Expected no. of desync oscillators}
    \includegraphics[width=\textwidth]{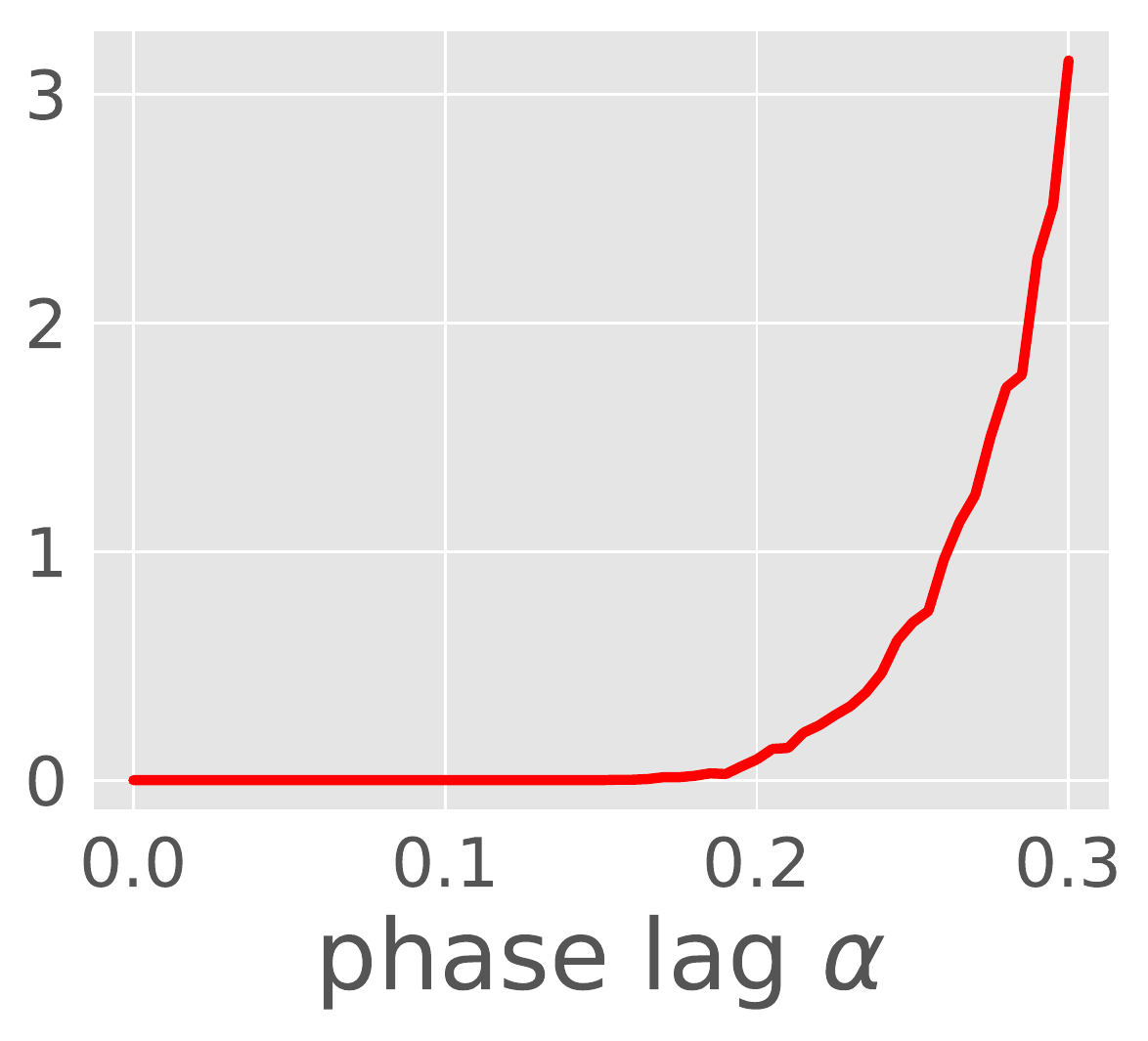}
  \end{subfigure}

 \label{fig:week1/northern hd 0}
\end{figure}

\end{document}